\documentclass{svjour3}
\smartqed
\usepackage{amsmath,amssymb}
\usepackage{graphicx}
\usepackage{epstopdf}
\usepackage{rotating}
\usepackage{tabularx}
\usepackage{lscape}
\usepackage{array}
\newcolumntype{L}[1]{>{\setlength{\parskip}{0pt}%
\setlength{\parindent}{0pt}%
\raggedright\arraybackslash}p{#1}}
\usepackage[square,comma,numbers,sort&compress]{natbib}

\usepackage{hyperref}
\usepackage{orcidlink}
\usepackage[english]{babel}

\usepackage{amsmath,amssymb,amsfonts}
\usepackage{xurl}
\usepackage{url}
\usepackage{hyperref}
\usepackage{orcidlink}

\usepackage{booktabs}
\usepackage{multirow}
\usepackage{array}
\usepackage{tabularx}
\usepackage{siunitx}
\usepackage{makecell}

\makeatletter
\renewcommand{\@biblabel}[1]{[#1]}
\makeatother

\journalname{European Physical Journal C}
\begin{document}

\title{Evaporation of Primordial Black Holes with Multimodal 
Mass and Extended Spin Distributions: Cosmological 
Imprints on the Effective Number of Relativistic Species}%

\titlerunning{Multimodal PBH Evaporation and the Effective Number of Relativistic Species}

\author{
T.~Toghrai\textsuperscript{*}\,\orcidlink{0000-0001-7142-0158},
A.~Daassou\,\orcidlink{0000-0001-9439-5047},
Y.~Ouchhaine\,\orcidlink{0009-0007-6764-9989},
H.~Laassiri\,\orcidlink{0009-0008-4772-1629},
R.~Benbrik\,\orcidlink{0000-0002-5159-0325}}

\authorrunning{Toghrai et al.}
\institute{
T.~Toghrai, A.~Daassou, Y.~Ouchhaine, H.~Laassiri, R.~Benbrik
\at
Energy, Environment, and Applications (LP2EA),
Polydisciplinary Faculty, Laboratory of Physics,
Cadi Ayyad University,
Sidi Bouzid, P.O. Box 4162, Safi, Morocco\\
\email{fssm.tahir@gmail.com}\\
  {\footnotesize $^{*}$Corresponding author}
}

\date{Received: 17 June 2026 / Accepted: 6 September 2026}
\maketitle

\begin{abstract}
Primordial black holes (PBHs) form through several distinct mechanisms, each imprinting a characteristic mass function that a single-mass treatment risks erasing.
Building on the \texttt{FRISBHEE} code, we introduce \texttt{FRISHBEE} (\url{https://doi.org/10.5281/zenodo.22699162}), implementing four mass functions -- log-normal, power-law, critical collapse, metric preheating -- in a multimodal framework, with spin modeled via Gaussian and Fishbach et al. profiles. Crucially, each channel's weight is derived from the primordial collapse probability rather than fitted, tying the mass function to the power-spectrum amplitude -- a concrete link between inflationary model-building and CMB observables.
Solving the Friedmann-Boltzmann equations for $M_{\rm PBH}^{\rm in}=10^{7}$~g in the pre-BBN window, we compute $\Delta N_{\rm eff}$ for five distributions, three spin configurations, and three weighting scenarios.
Extended mass functions enhance $\Delta N_{\rm eff}$ over the monochromatic benchmark by factors of $\sim\!1.03$ (critical collapse) to $\sim\!1.84$ (log-normal), preserving the hierarchy $LN > MM > PL > MP > CC$ across the mass window $\{10^{5},10^{7},10^{8}\}$~g.
Near-extremal spin ($a_{\star}=0.99$) and spin-2 dark radiation invert this trend: superradiant spin loss averages away faster in broad distributions than in a monochromatic population, letting the monochromatic approximation overestimate $\Delta N_{\rm eff}$.
Against CMB-S4 and Simons Observatory sensitivities ($\sigma\simeq 0.06$ and $0.05$), the monochromatic benchmark is undetectable for scalar dark radiation (sDR $=0$), while log-normal, power-law, and multimodal distributions reach $\sim\!1.6$--$1.7\sigma$; for spin-2 dark radiation the signal drops over an order of magnitude, with none detectable.
The formation channel thus governs the magnitude and observational accessibility of the Hawking signal, establishing $\Delta N_{\rm eff}$ as a discriminant between single- and multi-channel PBH formation scenarios.
\end{abstract}
\maketitle
\section{Introduction}\label{sec:introduction}

Primordial black holes (PBHs) are hypothetical compact objects predicted to form in the early Universe from the gravitational collapse of large-amplitude density perturbations re-entering the Hubble horizon during radiation domination~\cite{carr1974black, carr1975primordial, hawking1974black, hawking1975particle, carr2010new, carr2021constraints, sasaki2018primordial, carr2020primordial, ESCRIVA2024261}. Unlike astrophysical black holes, PBHs can span an enormous range of masses, from far below a gram to thousands of solar masses, set by the epoch and mechanism of formation. This breadth of possible masses is what makes PBHs cosmologically interesting: depending on their mass, PBHs have been proposed as some or all of the dark matter~\cite{cheek2022primordial, carr2010new, carr2021constraints}, as progenitors of a stochastic gravitational-wave background, and as participants in baryogenesis. PBHs light enough to have evaporated by today ($M\lesssim 10^{15}\,\mathrm{g}$) leave a complementary, indirect signature: their Hawking evaporation injects entropy and particle species into the primordial plasma, an effect potentially observable today as a shift in the effective number of relativistic species $N_{\rm eff}$~\cite{cheek2023evaporation, cheek2022primordial}. It is this channel -- accessible neither through direct detection nor through gravitational lensing, but through precision cosmology -- that we study in this work.

$\Delta N_{\rm eff}$ sourced by PBH evaporation has been studied by several groups. Lunardini and Perez-Gonz\'alez~\cite{lunardini2020dirac} established that $\Delta N_{\rm eff}$ cannot generically be large for a monochromatic PBH population, setting a quantitative ceiling on the achievable signal. Masina~\cite{masina2021dark} showed separately that large initial PBH spin can significantly enhance $N_{\rm eff}$ through superradiant amplification of Hawking emission from near-extremal Kerr black holes -- a result we revisit and extend to extended and multimodal mass distributions in Sec.~\ref{sec:spin_results}.

A subtlety central to the sign of $\Delta N_{\rm eff}$, discussed in detail by Refs.~\cite{hasegawa2019mev, carr2021constraints} and related literature, is the cosmic epoch of evaporation relative to neutrino decoupling ($T \lesssim 2$--$3\,\mathrm{MeV}$, i.e.\ $M_{\rm PBH} \gtrsim 10^{9}\,\mathrm{g}$). Evaporation before decoupling heats photons and neutrinos democratically and always yields $\Delta N_{\rm eff}>0$; evaporation at or after decoupling pits electromagnetic dilution of the photon-heated bath -- which lowers $N_{\rm eff}$ below its standard value of 3.044~\cite{bennett2020towards, bennett2021towards} -- against direct Hawking production of neutrinos, and can yield $\Delta N_{\rm eff}<0$~\cite{hasegawa2019mev, carr2021constraints, lunardini2020dirac}. We work exclusively in the pre-BBN window $M_{\rm PBH}^{\rm in}\in\{10^{5},10^{7},10^{8}\}\,\mathrm{g}$, where $T_{\rm evap}\gg T_{\rm dec}$ and this dilution is absent, so that $\Delta N_{\rm eff}>0$ throughout, as confirmed numerically in Sec.~\ref{sec:results}. Extended distributions with a large width, such as the log-normal with $\sigma=1$ centered at $M_c=10^{7}\,\mathrm{g}$, nonetheless develop a high-mass tail reaching toward $\sim10^{8}$--$10^{9}\,\mathrm{g}$, close enough to the BBN boundary to receive a partial dilution correction; a full treatment of this effect is deferred to future work. The treatment of the neutrino sector and this caveat are detailed in Sec.~\ref{sec:neff_sign}.

A central difficulty in PBH phenomenology is that the mass function is not uniquely fixed by theory: different formation mechanisms predict qualitatively different shapes. A localized enhancement of the primordial power spectrum, as in an ultra-slow-roll phase of inflation, produces a log-normal mass function~\cite{dolgov1993baryon, garcia1996density}; a scale-invariant spectrum collapsing during an era of constant equation of state $w$ produces a power-law mass function~\cite{carr1975primordial}; critical scaling near the collapse threshold produces a mass function with a low-mass power-law tail and an exponential cutoff at the horizon mass~\cite{niemeyer1998near}; and resonant amplification of perturbations during inflaton oscillations at the end of inflation (metric preheating) produces a numerically determined mass function peaked at a model-dependent scale~\cite{martin2020primordial, auclair2021primordial}. These four mechanisms are not mutually exclusive: in a realistic inflationary history several of them can operate at different epochs and scales, yet the great majority of PBH studies model the population as monochromatic, or at best as a single extended distribution at a time~\cite{cheek2023evaporation}.

PBH spin adds a further layer of model dependence. PBHs generically form with negligible angular momentum, but can acquire substantial spin through binary capture and mergers, particularly during matter-dominated epochs. Hierarchical mergers drive the spin distribution toward a universal asymptotic form peaked near $a_\star\approx0.7$~\cite{fishbach2017ligo}, essentially independently of the initial conditions -- a universality that lets us factorize the joint mass--spin distribution, as we do throughout this work.

Cheek et al.~\cite{cheek2023evaporation} advanced this picture considerably by implementing extended, single-channel mass and spin distributions in the public code FRISBHEE, showing that distribution effects alone can substantially change $\Delta N_{\rm eff}$ relative to the monochromatic approximation. What has not been addressed is the case where several formation channels operate \emph{simultaneously}, producing a genuinely multimodal PBH population -- the generic expectation once one allows for more than one phase of the post-inflationary Universe (ultra-slow-roll inflation, metric preheating, radiation-dominated critical collapse, and near-scale-invariant epochs) to source PBHs. Crucially, when several channels coexist, the relative weight of each is not a free modeling choice: it is fixed by the amplitude of the primordial power spectrum at each channel's characteristic scale, through the collapse probability~\cite{green2004new, musco2009primordial, musco2013primordial}. Multi-feature inflationary models with several transient phases of ultra-slow-roll or resonant particle production~\cite{garcia1996density, carr2010new} generically produce a hierarchy of power-spectrum peaks, each seeding a PBH sub-population with a calculable abundance. This is the physical picture our multimodal framework is designed to capture: it supplies the missing phenomenological interface between such inflationary models and $N_{\rm eff}$ measurements, with population fractions that are, in principle, derivable from the primordial power spectrum rather than fitted by hand. This is what makes the multimodal signal more than an incremental generalization of the extended-distribution case: it turns $\Delta N_{\rm eff}$ into a potential probe of the number and relative amplitude of features in the inflaton potential, accessible to CMB experiments already being built.

In this work we close this gap by modeling the global PBH population as a linear superposition of the four channels described above, with fractions derived self-consistently from the collapse probability of each channel (Sec.~\ref{sec:distributions}), and by examining the robustness of the resulting $\Delta N_{\rm eff}$ signal under three physically motivated weighting schemes. We implement this framework in \texttt{FRISHBEE}, our extension of \texttt{FRISBHEE}, and use it to compute $\Delta N_{\rm eff}$ -- the principal cosmological observable sourced by Hawking evaporation and the target of current and forthcoming CMB experiments. Relative to the existing literature~\cite{lunardini2020dirac, masina2021dark, chattopadhyay2025dark, creque2021resonant, hasegawa2019mev, carr2021constraints, cheek2023evaporation}, the novel contribution of this work is precisely this: a systematic treatment of \emph{multimodal} PBH mass distributions grounded in the primordial collapse probability, together with a quantification of the resulting enhancement of $\Delta N_{\rm eff}$, and of its detectability, across mass scales, weighting schemes, and spin configurations. A companion study of the PBH dark matter relic abundance across the same extended and multimodal distributions is left to forthcoming work; likewise, the BBN bound $M_{\rm PBH}^{\rm in}\lesssim10^{9}\,\mathrm{g}$ is used only to define the mass window studied here and is not itself computed.

This paper is organized as follows. Section~\ref{sec:distributions} reviews the four PBH mass distributions and introduces the multimodal mixture and its physically derived population fractions, together with the spin distributions considered. Section~\ref{sec:framework} presents the Friedmann--Boltzmann framework, the comoving phase-space formulation, and the Kerr Hawking-emission formalism used to evolve the PBH population. Section~\ref{sec:results} presents and discusses our numerical results for $\Delta N_{\rm eff}$, including their observational significance for CMB-S4 and the Simons Observatory. We conclude in Sec.~\ref{sec:conclusion}.

\section{PBH mass and spin distributions}\label{sec:distributions}
The cosmological imprints of PBH evaporation depend 
sensitively on the mass and spin distributions of the 
PBH population, rather than on a single characteristic 
mass and spin alone. We therefore begin by specifying 
the four mass functions considered in this work, each 
corresponding to a distinct physical formation channel, 
before introducing the multimodal mixture that forms 
the central new element of our framework. The spin 
distributions are discussed in 
Sec.~\ref{sec:spin_dist}. Throughout, we follow the 
parametrisation and fiducial parameter choices of 
Cheek et al.~\cite{cheek2023evaporation}, to which 
we refer the reader for further details on the 
numerical implementation of the individual 
distributions.

\subsection{Mass Distributions}
\subsubsection{Log-Normal (LN) }
Primordial black holes (PBHs) are formed in many inflationary scenarios when there is a localized enhancement in the primordial power spectrum of curvature perturbations, such as those produced during a brief ultra-slow-roll inflationary phase \cite{dolgov1993baryon} . Naturally, a mass function with a localized peak like that is well described by a log-normal distribution. The associated PBH mass function is expressed as
\begin{equation}
f_{\rm PBH}(M) = \frac{1}{\sqrt{2\pi}\,\sigma\, M}\,
\exp\!\left[-\frac{\left(\log^2(M/M_c)\right)}{2\sigma^2}\right],
\end{equation}
where $\sigma$  regulates the distribution's width and $M_c$ indicates its central (or characteristic) mass.

The scale at which the primordial power spectrum is enhanced determines the parameter $M_c$, which translates to the typical horizon mass at the moment when the relevant perturbations re-enter the Hubble horizon and collapse into PBHs. 
For small $\sigma$ values, the distribution is narrow and almost a monochromatic mass function, whereas for larger $\sigma$ values, the distribution is broader spanning several orders of magnitude in mass. The width $\sigma$ indicates the strength and duration of the enhancement of primordial fluctuations.

For phenomenological purposes, the quantity $M f_{\rm PBH}(M)$ is often helpful, as it shows the distribution of the total PBH mass density across different mass scales. Narrow log-normal distributions have a sharp peak around $M_c$, while wider distributions have a lot of mass both below and above $M_c$. Such extended mass functions directly shape the 
evaporation history and the resulting cosmological imprints, motivating the systematic study undertaken in this work.

\subsubsection{Power-Law (PL) }
A scale-invariant power spectrum of primordial perturbations collapsing in a universe dominated by a perfect fluid with constant equation-of-state parameter $w$ \cite{carr1975primordial} is another potential mechanism for the formation of primordial black holes (PBHs). 
Over a finite mass range, the resulting PBH mass function in this instance has a power-law form and can be expressed as
\begin{equation}
f_{\rm PBH}(M) \propto
\begin{cases}
M^{-\alpha}, & M_c \leq M \leq 10^{\sigma} M_c, \\
0, & \text{else},
\end{cases}
\end{equation}
where the exponent $\alpha$ is given by
\begin{equation}
\alpha = \frac{4w+2}{w+1}.
\end{equation}
The range of scales over which the primordial power spectrum is roughly scale invariant determines the mass interval $[M_c,\,10^{\sigma}M_c]$. 
When the Universe is not inflating, the equation-of-state parameter usually has values in the range $-1/3 < w \leq 1$, which results in $\alpha>1$ and a mass function dominated by lighter PBHs. 
The power-law distribution, in contrast to the log-normal case, continuously spans several decades in mass and does not show a preferred mass scale within the permitted interval.

The quantity $M f_{\rm PBH}(M)$ is again helpful for phenomenological applications in order to visualize the distribution of the PBH mass density across masses. 
The relative importance of PBH evaporation at various cosmic times can be greatly impacted by the distribution being dominated by either the low-mass end or the high-mass end of the permitted mass range, depending on the value of $\alpha$.

\subsubsection{Critical Collapse (CC)}
Primordial black hole (PBH) formation from primordial density fluctuations is more precisely described by applying critical scaling to gravitational collapse. 
Critical collapse results in an extended mass function with an upper cut-off around the horizon mass and a characteristic tail towards lower masses, a feature that seems to be generic in a broad class of inflationary models, whereas early treatments assumed that PBHs formed with a mass comparable to the horizon mass at formation.

The resulting PBH mass function can be conveniently 
parametrized following Niemeyer \& 
Jedamzik~\cite{niemeyer1998near} as:
\begin{equation}
f_{\rm PBH}(M) \propto M^{\gamma}\,
\exp\!\left[-\left(\frac{M}{M_c}\right)^{\delta}\right],
\end{equation}
where $M_c$ is the characteristic mass scale corresponding 
to the first peak of the distribution, and $\gamma = 1.25$, 
$\delta = 2.85$ are the critical scaling exponents 
adopted in this work and implemented in FRISHBEE.

This parametrization captures both the critical scaling behavior at low masses and the exponential suppression at the horizon scale, and has been shown to provide a good description of numerical collapse simulations~\cite{niemeyer1998near}.

\subsubsection{Metric Preheating (MP) }
In genuine particle-physics-driven inflationary models, the end of inflation is frequently succeeded by a phase of early matter dominance, propelled by the coherent oscillations of the inflaton field. 
During this phase, primordial disturbances produced at the end of inflation can be resonantly amplified and potentially collapse into primordial black holes (PBHs) prior to the reheating of the Universe. 
This formation process, commonly referred to as metric preheating, can therefore lead to the efficient creation of PBHs.

The PBH mass function does not admit a simple 
closed analytical form and must instead be 
determined from numerical simulations. 
This study utilizes the reference distribution delineated in the Appendix of Ref.\cite{cheek2023evaporation, martin2020primordial, auclair2021primordial}, characterized by a peak at a characteristic mass \(M_c^{\rm AV} \sim 10^{5.6}\,\mathrm{g}\). 
We assume that the overall form of the distribution stays the same when we adjust the reheating temperature in order to look at different characteristic mass scales. Then, we rescale the distribution in logarithmic mass as
\begin{equation}
f_{\rm PBH}(M) = f_{\rm PBH}^{\rm AV}\!\left(M \times \frac{M_c^{\rm AV}}{M_c}\right).
\end{equation}
This prescription allows us to explore a variety of different PBH mass scales in the metric-preheating scenario while preserving the numerically determined 
shape of the mass function. 

\subsubsection{Multimodal Distribution}\label{sec:multimodal}

In a realistic cosmological history, the post-inflationary Universe 
undergoes a succession of distinct phases, each of which can 
independently source a population of primordial black holes. During 
inflation, localized enhancements of the primordial power spectrum, 
as produced for instance by an ultra-slow-roll phase or by multi-field 
dynamics, generate PBHs whose mass function is well described by a 
log-normal distribution peaked at the horizon mass corresponding to 
the enhanced scale~\cite{dolgov1993baryon, garcia1996density}. At the end 
of inflation, the coherent oscillations of the inflaton field drive a 
phase of metric preheating during which resonantly amplified 
perturbations can collapse into PBHs before reheating is 
complete~\cite{martin2020primordial, auclair2021primordial}. Throughout the radiation-dominated era, 
the gravitational collapse of density perturbations near the formation 
threshold follows critical scaling, producing a characteristic 
low-mass tail described by the critical collapse mass 
function~\cite{niemeyer1998near}. Finally, over scales where the 
primordial power spectrum is approximately scale-invariant, the 
resulting PBH mass function follows a power-law~\cite{carr1975primordial}.

Crucially, these four mechanisms operate at \textit{different mass 
scales} and at \textit{different epochs} in cosmic history. Their 
respective PBH populations therefore coexist in the Universe without 
mutual interference, and their combined effect on cosmological 
observables cannot be captured by treating each channel in isolation. 
Several recent works have highlighted the possibility of such 
multi-component PBH populations: inflationary models with multiple 
features in the power spectrum naturally produce PBH populations at 
several distinct mass scales~\cite{garcia1996density, carr2010new, carr2021constraints}, and 
in scenarios where PBHs temporarily dominate the energy density before 
evaporating, multiple formation channels operating at different epochs 
can each contribute a distinct sub-population to the total PBH 
budget~\cite{cheek2023evaporation, martin2020primordial, auclair2021primordial}. This physical picture motivates 
the construction of a global multimodal PBH mass distribution as a 
superposition of the four individual components.

We model the global PBH mass distribution as a linear combination of the four separate populations:
\begin{equation}
    f_{\rm mix}(M) = \alpha_{LN}\, f_{LN}(M) + \alpha_{PL}\, f_{PL}(M) + \alpha_{CC}\, f_{CC}(M) + \alpha_{MP}\, f_{MP}(M),
\end{equation}
where $\alpha_i$ denotes the fractional energy-density contribution of each PBH sub-population at formation. The normalization condition
\begin{equation}
    \alpha_{LN} + \alpha_{PL} + \alpha_{CC} + \alpha_{MP} = 1
\end{equation}
ensures that $f_{\rm mix}(M)$ remains a legitimate probability density function.

\paragraph*{Physical derivation of the population fractions.}

The population fractions $\alpha_i$ are not free parameters. In any 
inflationary model, the fractional energy density of PBHs produced by 
channel $i$ at formation is governed by the primordial collapse 
probability~\cite{carr1975primordial, green2004new}:
\begin{equation}
\beta'_i \simeq \gamma_i \int_{\delta_c}^{+\infty} 
\mathcal{P}(\delta;\,k_i)\,d\delta,
\label{eq:betai}
\end{equation}
where $\delta_c \approx 0.45$ is the critical density contrast for 
collapse in the radiation-dominated 
era~\cite{musco2009primordial, musco2013primordial, harada2013threshold}, 
$\gamma_i \sim \mathcal{O}(0.2$--$0.8)$ is the 
formation efficiency of channel $i$ (encoding the fraction of the 
Hubble volume that collapses, see e.g.\ Ref.~\cite{carr1975primordial}), 
and $\mathcal{P}(\delta;\,k_i)$ is the probability distribution 
of the smoothed density contrast at the characteristic scale $k_i$ 
corresponding to formation-channel $i$.
For Gaussian primordial perturbations, 
$\mathcal{P}(\delta;\,k_i)$ is a Gaussian of variance
\begin{equation}
\sigma^2(k_i) = \int \frac{dk}{k}\,\mathcal{W}^2(k/k_i)\,
\mathcal{P}_{\mathcal{R}}(k),
\end{equation}
where $\mathcal{P}_{\mathcal{R}}(k)$ is the primordial curvature 
power spectrum and $\mathcal{W}$ is a smoothing window function, 
so that Eq.~(\ref{eq:betai}) reduces to~\cite{green2004new}:
\begin{equation}
\beta'_i \simeq \gamma_i\,\mathrm{erfc}\!\left(
\frac{\delta_c}{\sqrt{2}\,\sigma(k_i)}\right).
\label{eq:betai_gaussian}
\end{equation}
The population fractions are then naturally defined as the fraction 
of the total PBH energy density at formation contributed by each channel:
\begin{equation}
\boxed{\alpha_i \equiv \frac{\beta'_i}{\beta'_{\rm tot}}, 
\qquad \beta'_{\rm tot} = \sum_j \beta'_j.}
\label{eq:alphai}
\end{equation}
Equation~(\ref{eq:alphai}) has a transparent physical interpretation: 
$\alpha_i$ is determined by the amplitude of the primordial power 
spectrum $\mathcal{P}_{\mathcal{R}}(k_i)$ at the characteristic 
wavenumber of each formation channel, weighted by its formation 
efficiency $\gamma_i$. These quantities are calculable for any 
specific multi-feature inflationary model (see 
e.g.~\cite{garcia1996density, carr2010new, martin2020primordial}), 
and are in principle constrained by CMB and large-scale structure 
observations on the relevant scales. The multimodal framework 
introduced here therefore provides the direct phenomenological 
interface between inflationary model building and CMB observables.

\paragraph*{Benchmark weighting scenarios.}
Since a fully specified multi-feature inflationary model is 
beyond the scope of the present paper, we work instead with 
three physically motivated benchmark scenarios for the 
$\alpha_i$, each corresponding to a different limiting 
behaviour of the collapse probability:

\begin{enumerate}
\item \textbf{Equal Abundance (EA).} When all four formation channels 
    operate with equal formation efficiency ($\gamma_i = \gamma$) and 
    the primordial power spectrum is enhanced to the same amplitude 
    at each characteristic scale ($\sigma(k_i) = \sigma_\star$ 
    for all $i$), Eq.~(\ref{eq:betai_gaussian}) yields 
    $\beta'_i = \beta'_\star$ for all $i$, so that:
    \begin{equation}
        \alpha_i^{\rm EA} = \tfrac{1}{4}, 
        \qquad i \in \{\text{LN, PL, CC, MP}\}.
    \end{equation}
    This is the maximally agnostic prior, in which no formation 
    channel is preferred \emph{a priori}.

\item \textbf{Mean-Mass Weighted (MW).}  A second physically 
    motivated scenario arises in inflationary models where 
    the power-spectrum enhancement is proportional to the 
    horizon mass at each characteristic scale, as naturally 
    occurs in models with a nearly scale-invariant red tilt 
    superimposed on the enhanced 
    features~\cite{garcia1996density, carr2010new}. 
    In this case $\beta'_i \propto M_c^{(i)}$, and since 
    the mean mass $W_i$ of each distribution is proportional 
    to $M_c^{(i)}$ up to a shape-dependent numerical prefactor, 
    this reduces to:
    \begin{equation}
        \alpha_i^{\rm MW} \propto W_i, 
        \qquad \alpha_i^{\rm MW} = \frac{W_i}{W_{\rm total}}.
    \end{equation}
    This scenario is therefore not a mathematical convenience 
    but a specific, self-consistent prediction of a class of 
    inflationary models. The mean masses $W_i$ of each 
    distribution are computed analytically below 
    (Sec.~\ref{sec:meanmasses}), yielding the 
    explicit fractions listed in 
    Table~\ref{tab:weightingschemes}.

\item \textbf{Number-Density Weighted (NW).} In the opposite 
    limit of scenarios where smaller-mass PBHs form with 
    higher abundance, as expected for steep power-law spectra 
    or for collapse near the quantum gravity scale, the 
    formation efficiency per unit mass favours light PBHs, 
    so $\beta'_i \propto 1/M_c^{(i)}$, giving:
    \begin{equation}
        \alpha_i^{\rm NW} \propto W_i^{-1}, 
        \qquad \alpha_i^{\rm NW} 
        = \frac{W_i^{-1}}{W_{\rm total}^{-1}}.
    \end{equation}
    In this scenario the power-law sub-population, 
    which has the smallest mean mass ($W_{\rm PL} \approx 29.9\,\mathrm{g}$), 
    dominates the mixture.
\end{enumerate}

The resulting population fractions for all three scenarios 
are summarized in Table~\ref{tab:weightingschemes}.
The robustness of our cosmological results with respect 
to this choice is demonstrated in 
Table~\ref{tab:multimodal_robustness}.

\begin{table}[!htbp]
\centering
\footnotesize
\renewcommand{\arraystretch}{1.3}
\setlength{\tabcolsep}{5pt}
\begin{tabular}{lcccc}
\hline
\textbf{Scenario} & $\alpha_{LN}$ & $\alpha_{PL}$ & $\alpha_{CC}$ & $\alpha_{MP}$ \\
\hline
Equal Abundance (EA) & $0.2500$ & $0.2500$ & $0.2500$ & $0.2500$ \\
Mean-Mass Weighted (MW) & $0.6974$ & $1.26\times10^{-5}$ & $0.1342$ & $0.1684$ \\
Number-Density Weighted (NW) & $1.81\times10^{-5}$ & $0.9998$ & $9.43\times10^{-5}$ & $7.51\times10^{-5}$ \\
\hline
\end{tabular}
\caption{Population fractions $\alpha_i$ for the three physically motivated 
weighting scenarios considered in this work. 
The EA scenario corresponds to equal collapse probability across channels 
(flat power spectrum, equal efficiencies). The MW scenario corresponds to 
$\beta'_i \propto M_c^{(i)}$ (mildly red-tilted spectrum). 
The NW scenario corresponds to $\beta'_i \propto 1/M_c^{(i)}$ 
(steep blue-tilted spectrum favouring light PBHs). 
The physical motivation for each is discussed in 
Sec.~\ref{sec:multimodal}. The MW scenario corresponds to the multimodal curve 
displayed in Figs.~\ref{fig:a}--\ref{fig:c}.}
\label{tab:weightingschemes}
\end{table}

\paragraph*{Analytical computation of mean masses.}\label{sec:meanmasses}
The mean mass $W_i$ of each distribution is computed as follows. 
We add a normalization constant $C$ to the raw distribution to 
ensure it integrates to unity, then evaluate 
$W_i = \int M\,f_{\rm PBH}(M)\,dM$ over the relevant physical interval.

\paragraph{a- Log-Normal (LN):}
The Log-Normal distribution is naturally centered on a mass $M_c$. Its mean mass integrated from $0$ to $+\infty$ is:
\begin{equation}
    W_{LN} = M_c \exp\!\left(\frac{\sigma^2}{2}\right).
\end{equation}
Using the reference parameters~\cite{cheek2023evaporation} ($M_c = 10^6\,\mathrm{g}$, $\sigma = 1$) gives $W_{LN} \approx 1.65 \times 10^6\,\mathrm{g}$.

\paragraph{b- Power-Law (PL):}
With normalization constant
$C = (1-\alpha)\,(M_{\max}^{1-\alpha} - M_{\min}^{1-\alpha})^{-1}$
over $[M_{\min}, M_{\max}] = [M_c, M_c\,10^\sigma]$, the mean mass is:
\begin{equation}
    W_{PL} = \frac{1-\alpha}{2-\alpha}\,
    \frac{M_{\max}^{2-\alpha} - M_{\min}^{2-\alpha}}{M_{\max}^{1-\alpha} - M_{\min}^{1-\alpha}}.
\end{equation}
For a radiation-dominated era ($w = 1/3$, $\alpha = 2.5$) with 
$M_{\min} = 10\,\mathrm{g}$, $M_{\max} = 10^6\,\mathrm{g}$~\cite{cheek2023evaporation}, 
we obtain $W_{PL} \approx 29.9\,\mathrm{g}$.

\paragraph{c- Critical Collapse (CC):}
Using the Euler Gamma function $\Gamma$, the normalization constant and mean mass are:
\begin{equation}
    C = \frac{\delta}{M_c^{\gamma+1}\,\Gamma\!\left(\frac{\gamma+1}{\delta}\right)}, 
    \qquad
    W_{CC} = M_c\,\frac{\Gamma\!\left(\frac{\gamma+2}{\delta}\right)}
    {\Gamma\!\left(\frac{\gamma+1}{\delta}\right)}.
\end{equation}
With $\gamma = 1.25$, $\delta = 2.85$, and 
$M_c = 10^{5.6}\,\mathrm{g}$~\cite{cheek2023evaporation}, 
we get $W_{CC} \approx 3.17 \times 10^5\,\mathrm{g}$.

\paragraph{d- Metric Preheating (MP):}
Because the distribution is sharply peaked, its mean mass 
is well approximated by its target central mass:
\begin{equation}
    W_{MP} \approx M_c \approx 10^{5.6}\,\mathrm{g} 
    \approx 3.98 \times 10^5\,\mathrm{g}.
\end{equation}

\paragraph{e- MW population fractions:}
Under the MW scenario, the total mean mass is 
$W_{\rm total} \approx 2.36 \times 10^6\,\mathrm{g}$, 
giving:
\begin{itemize}
    \item \textbf{Log-Normal:} $\alpha_{LN} \approx 0.6974\ (69.74\%)$
    \item \textbf{Metric Preheating:} $\alpha_{MP} \approx 0.1684\ (16.84\%)$
    \item \textbf{Critical Collapse:} $\alpha_{CC} \approx 0.1342\ (13.42\%)$
    \item \textbf{Power-Law:} $\alpha_{PL} \approx 1.26\times10^{-5}\ (0.00126\%)$
\end{itemize}
Under the MW scenario, the mixture is dominated by 
the log-normal component because the LN distribution 
has by far the largest mean mass, reflecting a 
power-spectrum amplitude proportional to the horizon 
mass at each scale. The NW scenario is dominated 
instead by the power-law component ($\alpha_{PL} \approx 99.98\%$), 
as its extremely small mean mass ($W_{PL} \approx 29.9\,\mathrm{g}$) 
corresponds to the highest formation rate per unit mass. 
The EA scenario assigns equal weight to all four channels.

\subsection{Spin Distributions}\label{sec:spin_dist}
Primordial black holes formed from the collapse of 
primordial perturbations generically carry negligible 
angular momentum at formation \cite{deluca2019initial, chiba2017spin, mirbabayi2020spin, deluca2020evolution}, since the initial 
perturbations are nearly spherically symmetric. 
However, PBHs can acquire significant spin through 
subsequent binary capture and merger events, 
which are most efficient during matter-dominated 
epochs when the Hubble rate is low relative to 
the binary capture rate.

Following a sufficient number of hierarchical 
mergers, the PBH spin distribution converges toward 
a universal asymptotic form that is largely 
insensitive to both the individual PBH masses and 
the initial spin configuration of the merging 
binaries~\cite{fishbach2017ligo}. Numerical and 
analytical investigations consistently find that 
this universal distribution is peaked near 
$\langle a_\star \rangle \approx 0.7$, with a 
strongly suppressed tail at $a_\star \lesssim 0.4$. 
Although the PBH masses considered in this work 
are many orders of magnitude below those of the 
astrophysical black holes studied in merger 
simulations, the universality of the asymptotic 
spin distribution renders these findings applicable 
to the present context.

This universality motivates the factorization of 
the joint mass--spin distribution prior to 
evaporation as
\begin{equation}
f_{\rm PBH}(M, a_\star) = 
f_{\rm PBH}(M) \times f_{\rm PBH}^{a}(a_\star),
\end{equation}
which decouples the mass and spin degrees of freedom 
and is adopted throughout this work. Two spin 
distributions are considered: the universal 
merger-driven distribution of Fishbach 
et al.~\cite{fishbach2017ligo}, valid for 
$0.4 \leq a_\star \leq 0.94$, and a Gaussian 
profile centered at $a_\star = 0.7$ with width 
$\sigma_{a_\star}$, which provides a smooth 
parametric representation of populations with 
a preferred spin value. The sensitivity of our 
results to the choice of spin distribution is 
examined in Sec.~\ref{sec:spin_results}.
Explicitly, the Gaussian profile is a normal distribution 
centered at a mean spin $\bar a_\star$ (equal to $0.7$ for 
the generic case discussed above, and to $0.99$ for the 
near-extremal configuration analyzed in 
Sec.~\ref{sec:spin_results}) with width $\sigma_{a_\star}$, 
truncated to the physical range $a_\star \in [0,1]$ and 
renormalized accordingly:
\begin{equation}
f^{a}_{\rm PBH}(a_\star) = \frac{1}{\mathcal{N}}\,
\exp\!\left[-\frac{(a_\star-\bar a_\star)^2}{2\sigma_{a_\star}^2}\right],
\qquad a_\star \in [0,1],
\label{eq:gaussian_spin}
\end{equation}
with the normalization constant
\begin{equation}
\mathcal{N} = \int_0^1 
\exp\!\left[-\frac{(a_\star-\bar a_\star)^2}{2\sigma_{a_\star}^2}\right] 
da_\star.
\label{eq:gaussian_spin_norm}
\end{equation}
The merger-driven distribution of Fishbach 
et al.~\cite{fishbach2017ligo}, by contrast, does not 
admit a simple closed-form expression: it is constructed 
numerically from a population of successive equal-mass 
black-hole mergers with isotropically distributed spin 
orientations, and is provided as a tabulated distribution 
peaked at $a_\star \approx 0.7$ with support over 
$0.4 \lesssim a_\star \lesssim 0.94$. We implement this 
tabulated distribution directly in FRISHBEE, following 
Cheek et al.~\cite{cheek2023evaporation}, rather than an 
analytic fit.
\section{Evaporation of PBH distributions}\label{sec:framework}
We present a phase-space distribution function $f_{PBH}(M,a_{*},t)$ to characterize the evolution of an extended population of primordial black holes (PBHs)~\cite{cheek2022primordial, cheek2022redshift}. The PBH mass $M$, the dimensionless spin parameter $a_*$, and the cosmological time $t$ depend on this function.

By integrating this distribution function throughout the whole phase space, we can obtain the macroscopic quantities that describe the PBH population. The physical number density $n_{BH}(t)$ and the energy density $\rho_{BH}(t)$ of the PBHs are specifically defined as follows:
\begin{equation}
    n_{BH}(t) = \int_0^1 \int_0^\infty f_{PBH}(M, a_*, t) \, dM \, da_*
\end{equation}
\begin{equation}
    \rho_{BH}(t) = \int_0^1 \int_0^\infty M f_{PBH}(M, a_*, t) \, dM \, da_*
\end{equation}
The initial PBH abundance is parametrized by 
$\beta' \equiv \rho_{\rm PBH}(t_{\rm in})/
\rho_{\rm tot}(t_{\rm in})$, the fraction of the 
total energy density in PBHs at formation. 
For $\beta' \ll 1$, PBHs never dominate the energy 
budget prior to evaporation; for 
$\beta' \gtrsim \beta'_{\rm dom}$, a transient 
matter-dominated epoch precedes evaporation. 
Throughout this work, we adopt $\beta' = 10^{-3}$ 
as a representative value in the sub-domination 
regime, consistent with the fiducial choice of 
Cheek et al.~\cite{cheek2023evaporation}.

\subsection{Continuity and Friedmann-Boltzmann Equations}\label{sec:continuity}
If we assume that the total number of PBHs stays the same throughout cosmic history (which means we don't count any late-time PBH production or mergers), then the distribution function must follow the continuity (or Liouville) equation in the phase space $(M,a_{*})$, as shown in~\cite{cheek2022primordial, cheek2022redshift}. The equation is written in its most basic form as follows:
\begin{equation}
    \frac{\partial f_{PBH}}{\partial t} + \frac{\partial}{\partial M} \left( f_{PBH} \frac{dM}{dt} \right) + \frac{\partial}{\partial a_*} \left( f_{PBH} \frac{da_*}{dt} \right) + 3H f_{PBH} = 0
\end{equation}
where $H = \dot{a}/a$ is the Hubble parameter, which takes into account how the Universe's expansion makes things less dense. The mass and spin 
loss rates $dM/dt$ and $da_*/dt$ are determined by the Hawking emission 
spectra originally computed by Page~\cite{PhysRevD.13.198, page1976particle} for 
massless particles from Schwarzschild and Kerr black holes respectively, 
with the full Standard Model particle content treated following 
MacGibbon and Webber~\cite{macgibbon1990quark, macgibbon1991quark} and implemented in FRISHBEE. The explicit Kerr emission rates and the resulting spin-evolution equation $da_\star/dt$ are derived in Sec.~\ref{sec:kerr_emission}.

We derive the Friedmann-Boltzmann equations to follow the evolution of the universe. Taking the time derivative of the PBH energy density $\rho_{BH}(t)$ gives:
\begin{equation}
    \dot{\rho}_{BH} = \int_0^1 \int_0^\infty M \frac{\partial f_{PBH}}{\partial t} \, dM \, da_*
\end{equation}
Substituting the continuity equation to replace the time derivative of the distribution function $\frac{\partial f_{PBH}}{\partial t}$:
\begin{equation}
    \dot{\rho}_{BH} = \int_0^1 \int_0^\infty M \left[ -3H f_{PBH} - \frac{\partial}{\partial M} \left( f_{PBH} \frac{dM}{dt} \right) - \frac{\partial}{\partial a_*} \left( f_{PBH} \frac{da_*}{dt} \right) \right] \, dM \, da_*
\end{equation}
Each term is evaluated separately. The first term yields $-3H \rho_{BH}$ directly. Assuming that the boundary terms vanish at $a_* = 0$ and $a_* = 1$, the third term, which involves the derivative with respect to spin $a_*$, equals zero.

For the second term, we can directly find the integral of the evaporation rate by performing an integration by parts on the mass $M$ and safely assuming that the physical boundary terms go away as $M \rightarrow 0$ and $M \rightarrow \infty$. Combining these findings, the Friedmann-Boltzmann equation governing the macroscopic dynamics of the PBH energy density is expressed as:
\begin{equation}
    \dot{\rho}_{BH} + 3H\rho_{BH} = \int_{0}^{1} \int_{0}^{\infty} \frac{dM}{dt} f_{PBH}(M,a_{*},t) \, dM \, da_{*}
    \label{eq:friedmann_boltzmann_pbh}
\end{equation}
At the same time, the energy that black holes lose by Hawking evaporation is added to the Standard Model (SM) radiation bath. The coupled equation shows how the SM energy density, $\rho_{SM}$, changes over time:
\begin{equation}
    \dot{\rho}_{SM} + 4H\rho_{SM} = - \int_0^1 \int_0^\infty \frac{\epsilon_{SM}}{\epsilon} \frac{dM}{dt} f_{PBH}(M, a_*, t) \, dM \, da_*
\end{equation}
where $\epsilon_{SM}$ and $\epsilon$ stand for the SM evaporation function and the total evaporation function, respectively. The negative sign makes sure that energy is conserved because the PBH mass loss rate ($dM/dt < 0$) means that energy is being added to the plasma.

\subsection{Treatment of Neutrino Decoupling and the Sign of \texorpdfstring{$\Delta N_{\rm eff}$}{Delta N\_eff}}
\label{sec:neff_sign}

A critical ingredient of the Friedmann--Boltzmann system concerns 
the treatment of the neutrino sector and its coupling to the 
electromagnetic plasma. This is not merely a technical detail 
but determines the \emph{sign} and magnitude of $\Delta N_{\rm eff}$, 
as first discussed systematically in Refs.~\cite{hasegawa2019mev, carr2021constraints, lunardini2020dirac}.

When PBH evaporation takes place at temperatures 
$T_{\rm evap} \gg T_{\rm dec} \sim 2$--$3\,\mathrm{MeV}$, 
electrons, positrons, photons, and neutrinos all remain in 
thermal equilibrium. Hawking emission heats the entire 
relativistic plasma democratically, with the entropy injected 
distributed among all particle species according to the 
Standard Model degrees of freedom. In this regime, 
$N_{\rm eff}$ is enhanced by the additional energy 
deposited into the plasma, yielding $\Delta N_{\rm eff} > 0$.

The situation changes qualitatively when evaporation occurs 
near or after neutrino decoupling. In this case, two competing 
effects determine the sign of $\Delta N_{\rm eff}$:

\begin{enumerate}
\item \textbf{Electromagnetic dilution}: Hawking emission into 
photons and $e^\pm$ pairs reheats the electromagnetic bath 
selectively, increasing the photon temperature relative to 
the already-frozen neutrino temperature. This reduces 
$N_{\rm eff}$ below its standard value of $3.044$ 
through the same mechanism as entropy injection from 
$e^\pm$ annihilation, but driven by PBH evaporation 
rather than thermal physics.

\item \textbf{Direct neutrino production}: PBHs also emit 
neutrinos directly in the Hawking spectrum, whose emission 
rate is governed by the Standard Model particle content 
at the Hawking temperature. This direct injection into 
the neutrino sector counteracts the dilution effect 
and tends to increase $N_{\rm eff}$.
\end{enumerate}

The competition between these effects was analyzed in 
detail in Ref.~\cite{hasegawa2019mev, carr2021constraints}, which demonstrated 
that for $M_{\rm PBH} \gtrsim 10^9\,\mathrm{g}$, 
the electromagnetic dilution dominates the direct 
neutrino production, resulting in 
$\Delta N_{\rm eff} < 0$ (i.e., $N_{\rm eff} < 3.044$). 
Conversely, for $M_{\rm PBH} \ll 10^9\,\mathrm{g}$, 
evaporation occurs well before neutrino decoupling 
and the dilution mechanism is inoperative, 
giving $\Delta N_{\rm eff} > 0$.

The FRISBHEE framework adopted in this 
work~\cite{cheek2023evaporation} tracks the photon 
and neutrino energy densities separately through 
dedicated equations incorporating the neutrino 
decoupling transition, and therefore captures 
this competition self-consistently when it is 
relevant. For the mass window studied here, 
$M_{\rm PBH}^{\rm in} \in \{10^5, 10^7, 10^8\}\,\mathrm{g}$, 
all characteristic masses satisfy 
$T_{\rm evap} \gg T_{\rm dec}$, placing the 
entire pre-BBN window in the positive 
$\Delta N_{\rm eff}$ regime. Our numerical 
results (Tables~\ref{tab:dneff_a0}--\ref{tab:dneff_a08}) 
confirm $\Delta N_{\rm eff} > 0$ throughout.

We draw attention, however, to a subtlety specific 
to extended distributions. For a log-normal 
distribution with $\sigma = 1$ centered at 
$M_c = 10^7\,\mathrm{g}$, the $2\sigma$ upper 
tail extends to masses 
$M \sim M_c\,e^{2\sigma} \approx 7 \times 10^8\,\mathrm{g}$, 
approaching the BBN boundary 
$M_{\rm BBN} \sim 10^9\,\mathrm{g}$. PBHs in 
this tail evaporate at temperatures close to 
$T_{\rm dec}$ and receive a partial 
electromagnetic dilution correction. 
This correction is sub-dominant for 
$M_c = 10^7\,\mathrm{g}$ but would become 
increasingly significant for distributions 
centered at $M_c \sim 10^8\,\mathrm{g}$ 
or with larger $\sigma$. A systematic 
study of the sign change and partial 
cancellation effects for extended 
distributions near the 
$M_{\rm BBN} \sim 10^9\,\mathrm{g}$ boundary 
is an important direction for future work.

\subsection{Kerr Hawking Emission: Mass and Spin Loss Rates}
\label{sec:kerr_emission}

We now make explicit the Hawking emission rates that enter the continuity equation above, for a PBH of mass $M$ and dimensionless spin parameter $a_\star \equiv J/M^2 \in [0,1]$. The horizon radius and horizon angular velocity of the corresponding Kerr geometry are
\begin{equation}
r_+ = M\left(1+\sqrt{1-a_\star^2}\right), \qquad
\Omega_H = \frac{a_\star}{2 r_+},
\label{eq:kerr_horizon}
\end{equation}
and the Hawking temperature is
\begin{equation}
T_{\rm H} = \frac{\sqrt{1-a_\star^2}}{4\pi r_+}
= \frac{\sqrt{1-a_\star^2}}{4\pi M\left(1+\sqrt{1-a_\star^2}\right)},
\label{eq:kerr_temperature}
\end{equation}
which reduces to the familiar Schwarzschild value $T_{\rm H}=1/(8\pi M)$ for $a_\star \to 0$.

For each Standard Model species $i$ of spin $s_i$ and internal degeneracy $g_i$, the number of quanta emitted per unit time and unit energy $\omega$ in the partial wave $(\ell,m)$ follows the generalization of the Hawking spectrum to Kerr black holes~\cite{PhysRevD.13.198, page1976particle, macgibbon1990quark, macgibbon1991quark, arbey2020evolution}:
\begin{equation}
\frac{d^2 N_i}{dt\,d\omega} = \frac{g_i}{2\pi}\sum_{\ell = s_i}^{\infty}\sum_{m=-\ell}^{\ell}
\frac{\Gamma_{s_i \ell m}(\omega, M, a_\star)}
{\exp\!\left[\dfrac{\omega - m\Omega_H}{T_{\rm H}}\right] - (-1)^{2s_i}},
\label{eq:hawking_spectrum}
\end{equation}
where $\Gamma_{s_i\ell m}(\omega,M,a_\star)$ is the greybody (absorption) factor obtained from the Teukolsky radial equation with purely ingoing boundary conditions at the horizon, and $-(-1)^{2s_i}$ enforces Bose--Einstein ($-1$, integer $s_i$) or Fermi--Dirac ($+1$, half-integer $s_i$) statistics. As is generic for Kerr black holes, $\Gamma_{s_i\ell m}$ admits no closed-form expression for arbitrary $(\omega, M, a_\star)$; throughout this work these greybody factors are obtained by numerically integrating the Teukolsky radial equation for each spin channel, following the methodology of the public code BlackHawk~\cite{arbey2019blackhawk, arbey2020evolution}, and are pre-tabulated over the relevant $(\omega,\ell,m,a_\star)$ grid for the full Standard Model particle content within FRISHBEE, inheriting the tabulation scheme of FRISBHEE~\cite{cheek2023evaporation}. Equation~\eqref{eq:hawking_spectrum} is the resulting expression evaluated numerically in FRISHBEE.

The mass and angular momentum loss rates follow as the energy- and angular-momentum-weighted moments of Eq.~\eqref{eq:hawking_spectrum}:
\begin{align}
\frac{dM}{dt} &= -\sum_i \int_0^\infty \omega\,\frac{d^2 N_i}{dt\,d\omega}\,d\omega
= -\frac{1}{2\pi}\sum_i g_i \sum_{\ell,m} \int_0^\infty
\frac{\omega\, \Gamma_{s_i\ell m}(\omega,M,a_\star)}
{\exp\!\left[\dfrac{\omega-m\Omega_H}{T_{\rm H}}\right]-(-1)^{2s_i}}\,d\omega,
\label{eq:dMdt}\\[4pt]
\frac{dJ}{dt} &= -\sum_i \int_0^\infty m\,\frac{d^2 N_i}{dt\,d\omega}\,d\omega
= -\frac{1}{2\pi}\sum_i g_i \sum_{\ell,m} \int_0^\infty
\frac{m\,\Gamma_{s_i\ell m}(\omega,M,a_\star)}
{\exp\!\left[\dfrac{\omega-m\Omega_H}{T_{\rm H}}\right]-(-1)^{2s_i}}\,d\omega.
\label{eq:dJdt}
\end{align}
Since $a_\star \equiv J/M^2$, the two rates combine through the chain rule into the explicit spin-evolution equation that enters the continuity equation of Sec.~\ref{sec:continuity} and drives the numerical evolution of $a_\star(t)$ shown in Fig.~\ref{fig:Mono}:
\begin{equation}
\boxed{\ \frac{da_\star}{dt} = \frac{1}{M^2}\frac{dJ}{dt} - \frac{2a_\star}{M}\frac{dM}{dt}\ }
\label{eq:dastardt}
\end{equation}

\paragraph*{Superradiance and its scaling with the spin of the emitted quantum.}
For co-rotating modes with $0<\omega<m\Omega_H$, the exponent in Eqs.~\eqref{eq:hawking_spectrum}--\eqref{eq:dJdt} becomes negative and the greybody factor itself changes sign, $\Gamma_{s\ell m}<0$: incident radiation is amplified rather than absorbed, and both energy and angular momentum are extracted from the hole at an enhanced rate. This is the Hawking-radiation realization of the classical superradiant scattering discovered by Zel'dovich and analyzed by Press \& Teukolsky~\cite{press1972floating} and Starobinsky~\cite{starobinski1973amplification} \& Churilov~\cite{starobinskij1973amplification}. The maximum superradiant amplification factor for the dominant $\ell=m=s$ mode of an extremal Kerr black hole grows steeply with the spin $s$ of the emitted quantum: of order $0.3\%$ for a scalar, $4.4\%$ for a photon, and as much as $138\%$ for a graviton, while half-integer-spin fermions exhibit \emph{no} superradiance, since Pauli blocking caps their occupation number at unity~\cite{press1972floating,starobinskij1973amplification, brito2015superradiance}. This spin-scaling is the physical origin of the sDR-dependence of the spin-induced enhancement reported in Tables~\ref{tab:dneff_a099}--\ref{tab:dneff_a08} and quantified in Sec.~\ref{sec:spin_results} below.

A direct consequence of Eq.~\eqref{eq:dastardt} is that the characteristic spin-down timescale $\tau_{a_\star}\equiv|a_\star/\dot a_\star|$ is generically much shorter than the mass-evaporation timescale $\tau_M\equiv|M/\dot M|$ for a near-extremal PBH, because the angular-momentum-carrying superradiant channels dominate $dJ/dt$ while contributing only a fraction of $dM/dt$. This is the behaviour visible in the top-left panel of Fig.~\ref{fig:Mono}, where $a_\star(\xi)$ falls toward zero while $f(\xi)=M(\xi)/M_{\rm PBH}^{\rm in}$ remains close to unity: a near-extremal PBH loses most of its spin well before it has lost an appreciable fraction of its mass. This asymmetry between $\tau_{a_\star}$ and $\tau_M$, and its dependence on the mass distribution, underlies the quantitative results discussed in Sec.~\ref{sec:spin_results}.

\subsection{Change of Variables and Comoving Integration}
To solve these equations numerically, we need to find the integrals at every cosmic time $t$. However, the analytical expression of the distribution is only definitively established at the initial moment of PBH creation, \(t_{in}\). To circumvent this difficulty and enable 
efficient numerical integration, we change the variables according to the approach in~\cite{cheek2022primordial, cheek2022redshift} so that the present phase space parameters $(M, a_{*})$ return to their original values at creation:
\begin{equation}
    (t_{in}, M_{in}, a_{*}^{in}) \longrightarrow (t, M, a_*)
\end{equation}
The preservation of the infinitesimal comoving number of PBHs guarantees the subsequent essential invariance:
\begin{equation}
    a^3(t) f_{PBH}(M, a_*, t) \, dM \, da_* = a^3(t_{in}) f_{PBH}(M_{in}, a_{*}^{in}, t_{in}) \, dM_{in} \, da_{*}^{in} \equiv \mathcal{F}_{in} \, dM_{in} \, da_{*}^{in}
\end{equation}
We have defined $\mathcal{F}_{in}$ as the initial comoving distribution that does not change over time.

We introduce the rescaled comoving energy densities, $\tilde{\rho}_{BH} \equiv a^3 \rho_{BH}$ and $\tilde{\rho}_{SM} \equiv a^4 \rho_{SM}$, to make the differential system easier to work with. The time derivative of $\tilde{\rho}_{BH}$ is:
\begin{equation}
    \dot{\tilde{\rho}}_{BH} = \frac{d}{dt}(a^3 \rho_{BH}) = 3a^2\dot{a}\rho_{BH} + a^3\dot{\rho}_{BH} = a^3(\dot{\rho}_{BH} + 3H\rho_{BH})
\end{equation}
Recognizing the parenthesized term as the 
left-hand side of Eq.~\eqref{eq:friedmann_boltzmann_pbh} 
and using comoving number conservation, we obtain 
the integral form for the PBH comoving energy density:
\begin{equation}
    \dot{\tilde{\rho}}_{BH} = \int_0^1 \int_0^\infty \frac{dM}{dt} \mathcal{F}_{in} \, dM_{in} \, da_{*}^{in}
\end{equation}
By employing a similar mathematical approach for the Standard Model comoving energy density $\tilde{\rho}_{SM}$, and recognizing that the time derivative of $a^4\rho_{SM}$ results in a $4H\rho_{SM}$ term, we derive:
\begin{equation}
    \dot{\tilde{\rho}}_{SM} = -a(t) \int_0^1 \int_0^\infty \frac{\epsilon_{SM}}{\epsilon} \frac{dM}{dt} \mathcal{F}_{in} \, dM_{in} \, da_{*}^{in}
\end{equation}
The full derivation of the Jacobian and the 
proof of comoving number density conservation 
are presented in the appendices of 
Ref.~\cite{cheek2023evaporation}, to which 
we refer the reader for details.

\section{Numerical results and discussions}\label{sec:results}

\paragraph*{The FRISHBEE code.}
All numerical results presented in this work are 
computed with FRISHBEE, our extended implementation 
of the public FRISBHEE code of 
Cheek et al.~\cite{cheek2023evaporation}. 
FRISHBEE incorporates two additions with respect 
to the original framework: 
(i) dedicated mass distribution modules implementing 
the four formation-channel mass functions considered 
in this work (log-normal, power-law, critical 
collapse, and metric preheating) as normalized 
probability density functions over the relevant 
physical mass intervals; and 
(ii) a multimodal distribution module that constructs 
the mass-weighted linear superposition 
$f_{\rm mix}(M)$ from these four components, 
accepting arbitrary population fractions $\alpha_i$ 
as derived in Sec.~\ref{sec:distributions} for each 
of the three weighting scenarios (EA, MW, NW). 
The underlying Friedmann--Boltzmann solver, the 
comoving phase-space integration scheme, the 
Hawking emission rates, and the Standard Model 
particle content are inherited without modification 
from FRISBHEE~\cite{cheek2023evaporation, 
cheek2022primordial, cheek2022redshift}. 
As a validation, we have verified that FRISHBEE 
reproduces the monochromatic FRISBHEE results 
of Cheek et al.~\cite{cheek2023evaporation} 
to numerical precision for all spin values 
and dark radiation species considered, 
confirming the correctness of the extended 
implementation.
We further note that the enhancement ratio 
$\Delta N_{\rm eff}/\Delta N_{\rm eff}^{\rm mono}$ 
reported in all tables is independent of $\beta'$, 
since this parameter enters linearly in both 
$\Delta N_{\rm eff}$ and $\Delta N_{\rm eff}^{\rm mono}$ 
and cancels exactly in the ratio. The absolute 
values of $\Delta N_{\rm eff}$ scale linearly 
with $\beta'$ and are quoted for 
$\beta' = 10^{-3}$ throughout.

We fix $M_{\rm PBH}^{\rm in} = 10^7\,\mathrm{g}$ as the common 
characteristic mass for all distributions. At this mass scale, 
all PBHs evaporate well before Big Bang Nucleosynthesis 
($M_{\rm BBN} \sim 10^9\,\mathrm{g}$) and, crucially, 
well before neutrino decoupling 
($T_{\rm dec} \sim 2$--$3\,\mathrm{MeV}$, 
corresponding to $M_{\rm PBH} \sim 10^9\,\mathrm{g}$), 
ensuring that their Hawking radiation contributes 
positively to $\Delta N_{\rm eff}$ without the 
electromagnetic dilution effect discussed in 
Sec.~\ref{sec:neff_sign}. 
The choice of this mass window also ensures that 
our results are not affected by the potential 
sign reversal of $\Delta N_{\rm eff}$ identified 
by Refs.~\cite{hasegawa2019mev, carr2021constraints, lunardini2020dirac} 
for evaporation occurring at or after neutrino 
decoupling.

This choice allows a clean comparison of the shape effects of the five 
distributions on cosmological observables, following the 
rescaling prescription described in 
Sec.~\ref{sec:distributions}: each distribution is centred 
at $M_c = M_{\rm PBH}^{\rm in} = 10^7\,\mathrm{g}$, 
so that any difference in $\Delta N_{\rm eff}$ reflects 
the shape of the mass function rather than a difference 
in characteristic mass scale. The multimodal population fractions $\alpha_i$ 
are computed at the natural physical mass scales of each 
formation mechanism, as dictated by their respective fiducial 
parameters~\cite{cheek2023evaporation}. These fractions 
characterise the intrinsic composition of the global PBH 
population and are therefore independent of the 
characteristic mass scale adopted for the numerical 
comparison. Each distribution is rescaled to 
$M_c = M_{\rm PBH}^{\rm in} = 10^7\,\mathrm{g}$ in order 
to isolate the effect of the \textit{shape} of the mass 
function on cosmological observables, independently of 
the effect of the characteristic mass scale. Under this 
rescaling, the population fractions $\alpha_i$ of the 
multimodal distribution are held fixed at their 
theoretically determined values, so that the shape of 
the mixture and hence its cosmological imprint 
are fully determined by the physical composition of the 
global PBH population.

Throughout this section, figures are shown for sDR $= 2.0$ 
(spin-2 graviton dark radiation) as a representative case. 
The dependence on sDR is quantified in 
Tables~\ref{tab:dneff_a0}--\ref{tab:dneff_a08}, which 
demonstrate that the enhancement ratios 
$\Delta N_{\rm eff}/\Delta N_{\rm eff}^{\rm mono}$ are 
largely independent of the dark radiation spin content. 
We emphasize that this near-independence concerns the 
\emph{ratio} $\Delta N_{\rm eff}/\Delta N_{\rm eff}^{\rm mono}$; 
the \emph{absolute} value of $\Delta N_{\rm eff}$ instead 
varies by more than an order of magnitude with sDR 
(Table~\ref{tab:dneff_a0}), a distinction made explicit 
in the Observational implications discussion below.
The multimodal (MM) curve displayed in all figures 
corresponds to the mean-mass-weighted (MW) scenario, 
adopted as the fiducial choice throughout; 
the robustness of the enhancement under the equal-abundance 
(EA) and number-density-weighted (NW) scenarios is 
quantified in Table~\ref{tab:multimodal_robustness}.

To verify that our conclusions are not specific to the 
choice $M_{\rm PBH}^{\rm in} = 10^7\,\mathrm{g}$, 
Table~\ref{tab:stability} reports the enhancement ratio 
$\Delta N_{\rm eff}/\Delta N_{\rm eff}^{\rm mono}$ for 
three representative mass scales 
$M_{\rm PBH}^{\rm in} \in \{10^5, 10^7, 10^8\}\,\mathrm{g}$, 
all within the pre-BBN evaporation window 
$M_{\rm PBH}^{\rm in} \ll M_{\rm BBN} \sim 10^9\,\mathrm{g}$. 
The enhancement ratio varies across the pre-BBN mass 
window, with 
the largest variation observed for the log-normal 
($\sim 42\%$), multimodal ($\sim 40\%$), and power-law 
($\sim 38\%$) distributions, and the smallest for 
the critical collapse distribution ($\sim 6.5\%$). This variation reflects the sensitivity 
of the mass-weighted evaporation history to the 
characteristic PBH mass scale, while the hierarchy 
of enhancement factors (LN $>$ MM $>$ PL $>$ MP $>$ CC) remains preserved across all three 
mass scales considered, confirming that the shape of 
the mass distribution is the primary driver of the 
enhancement.

\begin{table}[!htbp]
\centering
\footnotesize
\setlength{\tabcolsep}{4pt}
\renewcommand{\arraystretch}{1.2}
\resizebox{\linewidth}{!}{%
\begin{tabular}{|c|c|
    S[table-format=1.4, round-precision=4]|
    S[table-format=1.4, round-precision=4]|
    S[table-format=1.4, round-precision=4]|
    S[table-format=2.2, round-precision=2]|}
\hline
\textbf{Distribution} & \textbf{sDR} 
 & {\makecell{$\mathbf{\Delta N_{\rm eff}/\Delta N_{\rm eff}^{\rm mono}}$ \\ $\mathbf{(10^5\ g)}$}}
 & {\makecell{$\mathbf{\Delta N_{\rm eff}/\Delta N_{\rm eff}^{\rm mono}}$ \\ $\mathbf{(10^7\ g)}$}}
 & {\makecell{$\mathbf{\Delta N_{\rm eff}/\Delta N_{\rm eff}^{\rm mono}}$ \\ $\mathbf{(10^8\ g)}$}}
 & \textbf{Variation (\%)} \\
\hline
\multirow{4}{*}{Log-Normal} &
0.0 & 1.0734153458940878 & 1.8518281021867358 & 1.1074317556266784 & 42.03400741941035 \\
\cline{2-6}
& 0.5 & 1.0734569413384856 & 1.8527148595597274 & 1.106681205500273 & 42.04572575257666 \\
\cline{2-6}
& 1.0 & 1.0730943559206380 & 1.8453392982657364 & 1.1126743806025992 & 41.85825151172339 \\
\cline{2-6}
& 2.0 & 1.0728260579998996 & 1.838826218222031 & 1.116406442913629 & 41.64948945206079 \\
\hline
\multirow{4}{*}{Power Law} &
0.0 & 1.0534840079803700 & 1.6984263558978743 & 1.102206809793296 & 37.98603091181262 \\
\cline{2-6}
& 0.5 & 1.0537647825085157 & 1.6992609477120835 & 1.1013287876755136 & 38.00843851155135 \\
\cline{2-6}
& 1.0 & 1.0515259530970826 & 1.6928016283806680 & 1.1082278493727844 & 37.89393952693138 \\
\cline{2-6}
& 2.0 & 1.0500914497828897 & 1.6876163973639515 & 1.112443490293555 & 37.78628086158867 \\
\hline
\multirow{4}{*}{Critical Collapse} &
0.0 & 1.0005951843807124 & 1.0401672320855677 & 1.0678938454418345 & 6.469294264536989 \\
\cline{2-6}
& 0.5 & 1.0006037890545139 & 1.0406228558013624 & 1.067932773407794 & 6.469491146289948 \\
\cline{2-6}
& 1.0 & 1.0005393695716690 & 1.0371937700317857 & 1.067509088678243 & 6.457405703330190 \\
\cline{2-6}
& 2.0 & 1.0005029673651586 & 1.0344881416715739 & 1.0670210306484276 & 6.430095120716734 \\
\hline
\multirow{4}{*}{Metric Preheating} &
0.0 & 1.0043470841359246 & 1.1764542048597050 & 1.1011673947061733 & 14.67256315752558 \\
\cline{2-6}
& 0.5 & 1.0044095730583580 & 1.1789615195445509 & 1.1003017218348088 & 14.83165426730484 \\
\cline{2-6}
& 1.0 & 1.0039408460232535 & 1.1614315323576632 & 1.107027373098199 & 13.56067810732548 \\
\cline{2-6}
& 2.0 & 1.003674731366622 & 1.1518225248950802 & 1.111032155887947 & 12.86227649151433 \\
\hline
\multirow{4}{*}{Multimodal (MW)} &
0.0 & 1.0686923794427030 & 1.7829150000674887 & 1.1038231228751407 & 40.05068668403646 \\
\cline{2-6}
& 0.5 & 1.0688299205126393 & 1.7840718260100592 & 1.1029831539075519 & 40.11751929966386 \\
\cline{2-6}
& 1.0 & 1.0676920732424635 & 1.7749416268672522 & 1.1096098115579480 & 39.96233293095603 \\
\cline{2-6}
& 2.0 & 1.0669141050379298 & 1.7679020636127267 & 1.11368099045245 & 39.71841029973197 \\
\hline
\end{tabular}
}
\caption{Enhancement ratio 
$\Delta N_{\rm eff}/\Delta N_{\rm eff}^{\rm mono}$ 
across three characteristic mass scales 
$M_{\rm PBH}^{\rm in} \in \{10^5, 10^7, 10^8\}\,\mathrm{g}$, 
all within the pre-BBN evaporation window 
$M_{\rm PBH}^{\rm in} \ll M_{\rm BBN} \sim 10^9\,\mathrm{g}$, 
for all five distributions and four dark radiation spins 
sDR, with $a^* = 0.0$. 
The last column Variation (\%) gives the maximum relative 
variation of the ratio across the three mass values, 
defined as 
$100 \times (\max - \min)/\text{Ratio}_{10^7\,\mathrm{g}}$. 
Although the absolute value of the ratio depends on 
$M_{\rm PBH}^{\rm in}$, the hierarchy LN $>$ MM $>$ PL $>$ MP $>$ CC 
is preserved across all three mass scales, confirming 
that the shape of the mass distribution is the primary 
determinant of the cosmological imprint of $\Delta N_{\rm eff}$. All values are quoted to four significant figures, 
consistent with the numerical integration tolerance 
of FRISHBEE ($\text{rtol} = \text{atol} = 10^{-5}$).}
\label{tab:stability}
\end{table}

\subsection{Effect of Mass Distributions on \texorpdfstring{$\Delta N_{\rm eff}$}{Delta N\_eff}}
\begin{figure}[!htbp]
   \centering
    \includegraphics[width=\linewidth]{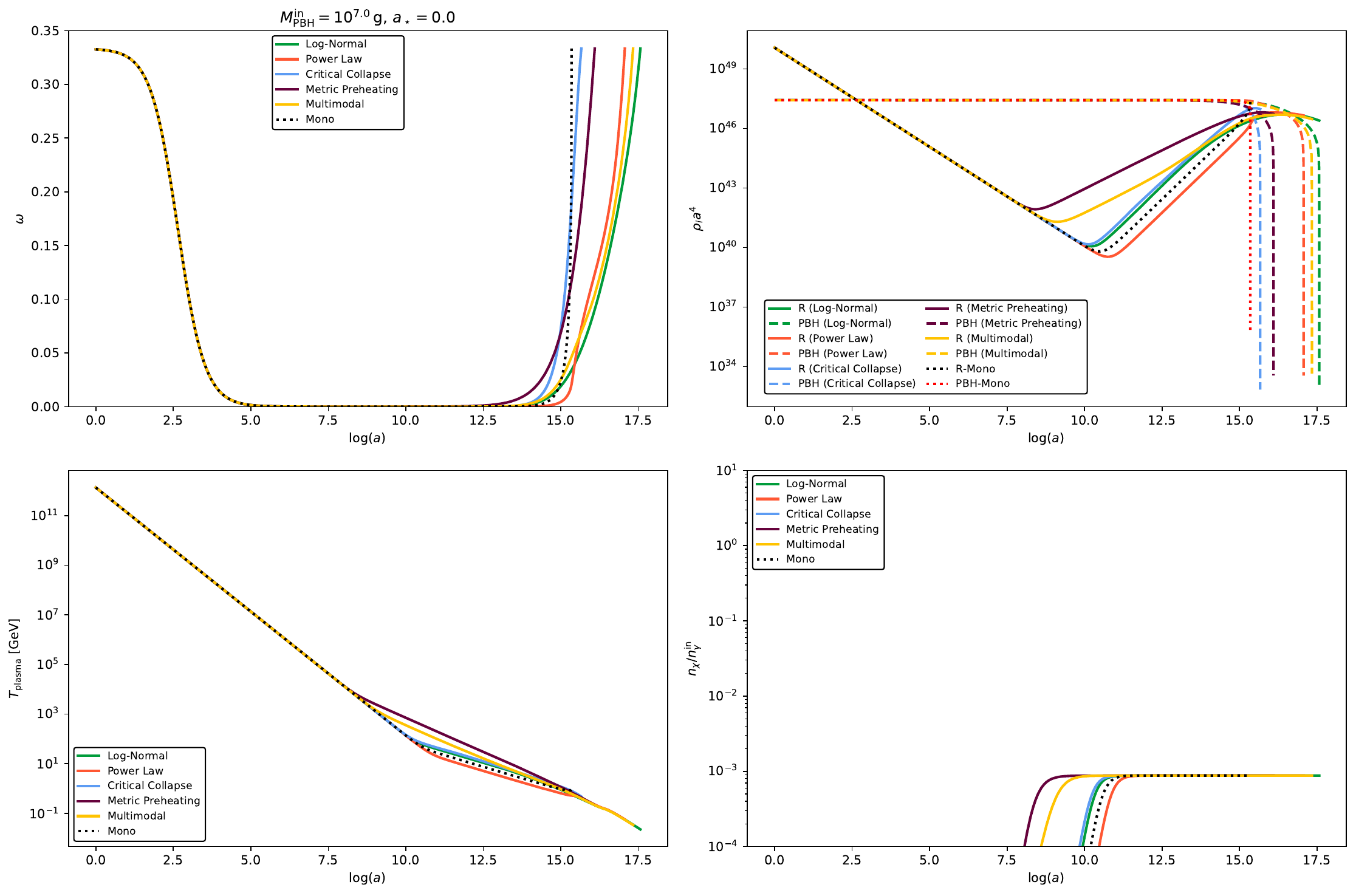}
    \caption{Evolution of the cosmological observables for all five 
PBH mass distributions considered in this work, namely log-normal (LN), 
power-law (PL), critical collapse (CC), metric preheating (MP), 
and multimodal (MM, here shown for the mean-mass-weighted scenario MW; 
see Table~\ref{tab:multimodal_robustness} for the EA and NW variants), 
compared to the monochromatic benchmark 
(dotted black), for $M_{\rm PBH}^{\rm in} = 10^7\,\mathrm{g}$, 
$a_\star = 0.0$, and sDR $= 2.0$ (spin-2 dark radiation). 
Results for other sDR values are summarized in 
Tables~\ref{tab:dneff_a0}--\ref{tab:dneff_a08}.
\textit{Top left:} equation-of-state parameter $w$ vs 
$\log(a)$, showing the transition from matter- to 
radiation-domination at PBH evaporation.
\textit{Top right:} comoving SM radiation (solid) and PBH 
(dashed) energy densities $\rho_i a^4$ vs $\log(a)$ for 
each distribution.
\textit{Bottom left:} plasma temperature $T_{\rm plasma}$ 
vs $\log(a)$.
\textit{Bottom right:} dark radiation number density ratio 
$n_\chi/n_\gamma^{\rm in}$ vs $\log(a)$, illustrating the 
distribution-dependent dark radiation production from 
Hawking evaporation.}
    \label{fig:a}
\end{figure}

We begin by examining the dependence of $\Delta N_{\rm eff}$ on the 
choice of PBH mass distribution for a fixed initial mass $M_{\rm PBH}^{\rm in} = 10^7\,\mathrm{g}$ and vanishing spin $a^* = 0.0$. 
The results are summarized in Table~\ref{tab:dneff_a0} and 
illustrated in Fig.~\ref{fig:a}.

\begin{table}[!htbp]
\centering
\footnotesize
\setlength{\tabcolsep}{4pt}
\renewcommand{\arraystretch}{1.2}
\resizebox{\linewidth}{!}{%
\begin{tabular}{|c|c|c|
    S[table-format=1.6, round-precision=6]|
    S[table-format=1.6, round-precision=6]|
    S[table-format=1.4, round-precision=4]|
    S[table-format=2.2, round-precision=2]|}
\hline
\textbf{Mass Distribution Type} 
 & \textbf{Free Parameters} & \textbf{sDR} 
 & $\mathbf{\Delta N_{\text{eff}}}$ 
 & $\mathbf{\Delta N_{\text{eff}}^{\text{mono}}}$ 
 & \textbf{Ratio} & \textbf{Diff (\%)} \\
\hline
\multirow{4}{*}{Log-Normal} & \multirow{4}{*}{$\sigma = 1.0$} & 
0.0 & 0.10470594152921969 & 0.05654193356585172 & 1.8518281021867358 & 85.18281021867358 \\
\cline{3-7}
& & 0.5 & 0.11344495388236042 & 0.061231739626311996 & 1.8527148595597274 & 85.27148595597274 \\
\cline{3-7}
& & 1.0 & 0.04588469595146146 & 0.024865181159141974 & 1.8453392982657364 & 84.53392982657365 \\
\cline{3-7}
& & 2.0 & 0.0052715005284923455 & 0.0028667747263193704 & 1.838826218222031 & 83.88262182220309 \\
\hline
\multirow{4}{*}{Power Law} & \multirow{4}{*}{$\sigma=1.0,\ \alpha=2.5$} & 
0.0 & 0.09603231018166924 & 0.05654193356585172 & 1.6984263558978743 & 69.84263558978743 \\
\cline{3-7}
& & 0.5 & 0.10404870390746646 & 0.061231739626311996 & 1.6992609477120835 & 69.92609477120835 \\
\cline{3-7}
& & 1.0 & 0.04209181915617584 & 0.024865181159141974 & 1.6928016283806680 & 69.28016283806680 \\
\cline{3-7}
& & 2.0 & 0.004838016035685124 & 0.0028667747263193704 & 1.6876163973639515 & 68.76163973639514 \\
\hline
\multirow{4}{*}{Critical Collapse} & \multirow{4}{*} {---} & 
0.0 & 0.05881306653395804 & 0.05654193356585172 & 1.0401672320855677 & 4.016723208556775 \\
\cline{3-7}
& & 0.5 & 0.06371914775561824 & 0.061231739626311996 & 1.0406228558013624 & 4.062285580136244 \\
\cline{3-7}
& & 1.0 & 0.025790010988973794 & 0.024865181159141974 & 1.0371937700317857 & 3.7193770031785807 \\
\cline{3-7}
& & 2.0 & 0.00296564445922116 & 0.0028667747263193704 & 1.0344881416715739 & 3.4488141671573826 \\
\hline
\multirow{4}{*}{Metric Preheating} & \multirow{4}{*} {---} & 
0.0 & 0.06651899549444436 & 0.05654193356585172 & 1.1764542048597050 & 17.645420485970508 \\
\cline{3-7}
& & 0.5 & 0.07218986479419308 & 0.061231739626311996 & 1.1789615195445509 & 17.896151954455085 \\
\cline{3-7}
& & 1.0 & 0.02887920545601316 & 0.024865181159141974 & 1.1614315323576632 & 16.14315323576632 \\
\cline{3-7}
& & 2.0 & 0.00330201570357458 & 0.0028667747263193704 & 1.1518225248950802 & 15.182252489508022 \\
\hline
\multirow{4}{*}{Multimodal (MW)} & \multirow{4}{*}{\shortstack{$\sigma=1.0,\ \alpha=2.5$\\ 
See Sec.~\ref{sec:multimodal}}} &
0.0 & 0.10080946148737646 & 0.05654193356585172 & 1.7829150000674887 & 78.29150000674886 \\
\cline{3-7}
& & 0.5 & 0.10924182152488694 & 0.061231739626311996 & 1.7840718260100592 & 78.40718260100591 \\
\cline{3-7}
& & 1.0 & 0.04413424509895640 & 0.024865181159141974 & 1.7749416268672522 & 77.49416268672522 \\
\cline{3-7}
& & 2.0 & 0.005068176954572825 & 0.0028667747263193704 & 1.7679020636127267 & 76.79020636127268 \\
\hline
\end{tabular}
}
\caption{Excess effective number of relativistic species $\Delta N_{\rm eff}$ 
for each mass distribution compared to the monochromatic benchmark, 
with $\mathbf{M_{\rm PBH}^{\rm in}}=10^7\,\mathrm{g}$ 
and $\mathbf{a^{\ast}}=0.0$.
The column ``Free Parameters'' lists the 
parameters varied in each distribution. 
For the critical collapse and metric preheating 
distributions, no free parameters are varied 
(indicated by ---): their shapes are fully 
determined by the physics of gravitational 
collapse~\cite{niemeyer1998near} and metric 
preheating~\cite{cheek2023evaporation, 
martin2020primordial, auclair2021primordial}, 
respectively. The multimodal result labelled (MW) uses 
the mean-mass-weighted population fractions of 
Sec.~\ref{sec:multimodal}; robustness across 
weighting schemes is demonstrated in 
Table~\ref{tab:multimodal_robustness}.
Here sDR denotes the spin of the dark radiation component, 
Ratio $= \Delta N_{\rm eff}/\Delta N_{\rm eff}^{\rm mono}$, 
and Diff $= 100\times(\Delta N_{\rm eff} - \Delta N_{\rm eff}^{\rm mono})/
\Delta N_{\rm eff}^{\rm mono}$. All values are quoted to four significant figures, 
consistent with the numerical integration tolerance of FRISHBEE ($\text{rtol} = \text{atol} = 10^{-5}$).}
\label{tab:dneff_a0}
\end{table}

The most striking result is that all extended distributions 
systematically enhance $\Delta N_{\rm eff}$ relative to the 
monochromatic benchmark, across all values of the dark radiation 
spin parameter sDR considered. This confirms that the monochromatic 
approximation generically underestimates the contribution of PBH 
evaporation to $N_{\rm eff}$, with the magnitude of the 
underestimation depending strongly on the formation channel.
These positive enhancements are consistent with the expectation 
that, in the pre-BBN mass window where $T_{\rm evap} \gg T_{\rm dec}$, 
the electromagnetic dilution of $N_{\rm eff}$ identified by 
Refs.~\cite{hasegawa2019mev, carr2021constraints, lunardini2020dirac} is absent and extended 
distributions enhance $\Delta N_{\rm eff}$ purely through 
their broader evaporation histories.

Among the four individual distributions, the log-normal (LN) 
distribution produces the largest enhancement, with 
$\Delta N_{\rm eff}/\Delta N_{\rm eff}^{\rm mono} \approx 1.84$ 
corresponding to an excess of $\sim 84\%$ over the monochromatic 
case. This can be understood physically: the log-normal distribution, 
with $\sigma = 1$, has significant weight at masses both below and 
above $M_{\rm PBH}^{\rm in}$, including a population of lighter PBHs 
that evaporate earlier and inject more energy per unit mass into the 
radiation bath at higher temperatures, when $g_*(T)$ is larger. This 
enhances the effective contribution to $\Delta N_{\rm eff}$ compared 
to a population of identical-mass PBHs.

The power-law (PL) distribution yields an intermediate enhancement 
of $\sim 1.69$ ($\sim 69\%$ excess), consistent with its broad 
support toward low masses governed by the exponent $\alpha = 2.5$. 
The steepness of the power-law ($\alpha > 2$) ensures that the 
distribution is dominated by light PBHs near $M_{\min}$, which 
evaporate efficiently and contribute substantially to 
$\Delta N_{\rm eff}$.

The metric preheating (MP) distribution produces a modest enhancement 
of $\sim 1.15$ ($\sim 15\%$ excess). Its relatively peaked shape 
around $M_c^{\rm AV} \sim 10^{5.6}\,\mathrm{g}$ limits the 
dispersion in evaporation times, resulting in a more 
monochromatic-like behaviour compared to the broader distributions.

The critical collapse (CC) distribution yields the smallest 
enhancement, $\sim 1.03$ ($\sim 3\%$ excess), approaching the 
monochromatic limit. This is consistent with the exponential 
suppression of the CC mass function at large masses and its 
relatively narrow effective width, which concentrates the PBH 
population near $M_{\rm PBH}^{\rm in}$.

The multimodal (MM) distribution under the MW scenario, whose 
population fractions are derived self-consistently from the 
mean-mass-weighted collapse probability as described in 
Sec.~\ref{sec:multimodal}, yields 
$\Delta N_{\rm eff}/\Delta N_{\rm eff}^{\rm mono} \approx 1.77$ 
($\sim 77\%$ excess), slightly lower than the log-normal case 
but comparable to the power-law distribution. This result reflects 
the dominant contribution of the log-normal sub-population 
($\alpha_{LN} \approx 69.74\%$ in the MW scenario) to the mixture, 
partially offset by the subdominant contributions from CC ($\approx 13.42\%$) 
and MP ($\approx 16.84\%$), which have smaller enhancements.

Notably, the ratio $\Delta N_{\rm eff}/\Delta N_{\rm eff}^{\rm mono}$ 
is remarkably stable across the four values of sDR considered 
($0.0, 0.5, 1.0, 2.0$) for each distribution. This indicates 
that the enhancement factor is primarily determined by the shape 
of the mass distribution rather than by the dark radiation spin 
content, and that our conclusions are robust with respect to 
assumptions about the dark radiation sector.

\subsubsection{Robustness with Respect to Weighting Scheme}

A key question is whether the multimodal signal depends 
sensitively on the choice of population fractions $\alpha_i$. 
Table~\ref{tab:multimodal_robustness} reports the multimodal 
enhancement ratio for all three physically motivated 
weighting scenarios introduced in Sec.~\ref{sec:multimodal}, 
at $M_{\rm PBH}^{\rm in} = 10^7\,\mathrm{g}$, $a^* = 0.0$, 
and sDR $= 2.0$.

\begin{table}[!htbp]
\centering
\footnotesize
\renewcommand{\arraystretch}{1.3}
\setlength{\tabcolsep}{5pt}
\resizebox{\linewidth}{!}{%
\begin{tabular}{|l|c|c|c|c|
    S[table-format=1.4]|
    S[table-format=1.4]|
    S[table-format=2.2]|}
\hline
\textbf{Weighting scenario} 
  & $\alpha_{LN}$ & $\alpha_{PL}$ & $\alpha_{CC}$ & $\alpha_{MP}$ 
  & {$\Delta N_{\rm eff}^{\rm MM}$} 
  & {Ratio} 
  & {Diff (\%)} \\
\hline
Equal Abundance (EA) 
  & $0.2500$ & $0.2500$ & $0.2500$ & $0.2500$ 
  & 0.0041 & 1.4282 & 42.82 \\
\hline
Mean-Mass Weighted (MW) 
  & $0.6974$ & $1.26\!\times\!10^{-5}$ & $0.1342$ & $0.1684$ 
  & 0.005068 & 1.7679 & 76.79 \\
\hline
Number-Density Weighted (NW) 
  & $1.81\!\times\!10^{-5}$ & $0.9998$ & $9.43\!\times\!10^{-5}$ & $7.51\!\times\!10^{-5}$ 
  & 0.004838 & 1.6876 & 68.76 \\
\hline
\textit{Monochromatic (reference)} & --- & --- & --- & --- 
  & \textit{0.002867} & \textit{1.0000} & \textit{---} \\
\hline
\end{tabular}
}
\caption{Multimodal enhancement ratio 
$\Delta N_{\rm eff}^{\rm MM}/\Delta N_{\rm eff}^{\rm mono}$ 
for the three physically motivated weighting scenarios 
of Sec.~\ref{sec:multimodal} (EA = equal abundance, 
MW = mean-mass weighted, NW = number-density weighted), 
at $M_{\rm PBH}^{\rm in} = 10^7\,\mathrm{g}$, $a^* = 0.0$, 
and sDR $= 2.0$. 
The EA $\Delta N_{\rm eff}$ value is computed as the 
linear combination 
$\tfrac{1}{4}(\Delta N_{\rm eff}^{LN} + \Delta N_{\rm eff}^{PL} 
+ \Delta N_{\rm eff}^{CC} + \Delta N_{\rm eff}^{MP})$; 
the NW value is dominated by the PL component 
($\alpha_{PL} \approx 99.98\%$). 
In all three scenarios, the multimodal signal exceeds 
the monochromatic benchmark by at least $\sim 43\%$, 
demonstrating that the enhancement is a robust prediction 
of multi-channel PBH formation independently of 
the assumed weighting scheme.}
\label{tab:multimodal_robustness}
\end{table}

The results of Table~\ref{tab:multimodal_robustness} 
demonstrate that in all three physically motivated scenarios, 
the multimodal $\Delta N_{\rm eff}$ exceeds the monochromatic 
benchmark by at least $\sim 43\%$ (EA scenario). The enhancement 
reaches $\sim 77\%$ in the MW scenario (dominated by the 
log-normal sub-population) and $\sim 69\%$ in the NW scenario 
(dominated by the power-law sub-population). The fact that 
the monochromatic approximation underestimates $\Delta N_{\rm eff}$ 
in all three cases, and by a substantial factor in each, 
confirms that the multimodal enhancement is not an artifact of 
any particular weighting choice: it is a genuine, robust 
prediction of multi-channel PBH formation scenarios.

\paragraph*{Observational implications.}
The absolute values of $\Delta N_{\rm eff}$ reported 
in Table~\ref{tab:dneff_a0} can be directly compared 
to the projected sensitivities of upcoming CMB 
experiments. CMB-S4~\cite{abazajian2016cmb} and the Simons 
Observatory~\cite{ade2019simons} are expected to reach 
$1\sigma$ sensitivities of $\sigma(\Delta N_{\rm eff}) 
\simeq 0.06$ and $\simeq 0.05$, respectively, on 
the excess number of relativistic species.

Because the \emph{absolute} value of $\Delta N_{\rm eff}$ 
depends strongly on the spin sDR of the dark radiation 
species, even though the \emph{enhancement ratio} 
$\Delta N_{\rm eff}/\Delta N_{\rm eff}^{\rm mono}$ does 
not (Table~\ref{tab:dneff_a0}), we first discuss 
detectability for scalar dark radiation (sDR $=0$), 
which yields the largest absolute $\Delta N_{\rm eff}$ 
among the sDR values considered and therefore the most 
favourable detection prospects; the sDR $=2$ case shown 
in Figs.~\ref{fig:a}--\ref{fig:c} is addressed separately 
below.

For the monochromatic benchmark at 
$M_{\rm PBH}^{\rm in} = 10^7\,\mathrm{g}$ and sDR $=0$, 
$\Delta N_{\rm eff}^{\rm mono} \simeq 0.057$ falls 
below the CMB-S4 detection threshold, rendering the 
monochromatic PBH contribution effectively invisible 
to next-generation CMB experiments at this mass scale. 
This is consistent with the general finding of 
Ref.~\cite{lunardini2020dirac} that $\Delta N_{\rm eff}$ 
from PBH evaporation cannot generically be large 
for monochromatic populations.
By contrast, the extended log-normal, power-law, and 
multimodal distributions yield 
$\Delta N_{\rm eff} \simeq 0.096$--$0.105$ (sDR $=0$), 
exceeding the CMB-S4 threshold at the $\sim 1.6$--$1.7\sigma$ 
level, while the metric preheating distribution 
marginally crosses the detection threshold with 
$\Delta N_{\rm eff} \simeq 0.067$. Only the critical 
collapse distribution, with its narrow effective 
width, remains below the experimental sensitivity 
with $\Delta N_{\rm eff} \simeq 0.059$.

We stress that this detectability window is specific 
to sDR $=0$. For sDR $=2$ (the spin-2 graviton case 
shown throughout the figures of this section), the 
absolute values of $\Delta N_{\rm eff}$ are suppressed 
by more than an order of magnitude relative to sDR $=0$ 
(Table~\ref{tab:dneff_a0}): e.g.\ 
$\Delta N_{\rm eff}^{\rm LN} \simeq 0.0053$ and 
$\Delta N_{\rm eff}^{\rm mono} \simeq 0.0029$ for 
$M_{\rm PBH}^{\rm in} = 10^7\,\mathrm{g}$, both far 
below $\sigma(\Delta N_{\rm eff})$ for CMB-S4 and the 
Simons Observatory. None of the distributions considered, 
extended or monochromatic, are therefore detectable for 
sDR $=2$ at this mass scale.

This comparison reveals a qualitative consequence of 
the monochromatic approximation that goes beyond a 
mere quantitative correction: for PBH populations 
described by realistic extended mass functions \emph{and 
for a scalar dark radiation species}, the 
Hawking evaporation signal in $\Delta N_{\rm eff}$ 
transitions from undetectable to potentially 
observable with CMB-S4 and the Simons Observatory. 
The formation channel therefore determines not only 
the magnitude of the cosmological imprint but also 
its observational accessibility, and this accessibility 
is itself contingent on the spin of the dark radiation 
species produced. These results 
motivate a systematic confrontation of multimodal 
PBH population models with forthcoming CMB data.

Because this detectability threshold is set by the shape of the mass function rather than by its overall normalization, a future measurement of $\Delta N_{\rm eff}$ at the sensitivity of CMB-S4 or the Simons Observatory would already carry diagnostic power over the PBH formation channel: a value near the extended-distribution predictions reported here would disfavor a purely monochromatic or critical-collapse-dominated PBH population at $M_{\rm PBH}^{\rm in}\sim10^{7}\,\mathrm{g}$, while a non-detection at the projected sensitivity would instead constrain the log-normal, power-law, and multimodal scenarios examined in this work. In this sense $\Delta N_{\rm eff}$ functions not only as an existence bound on PBHs but as a probe of the shape of their mass function.
\subsection{Effect of Spin on \texorpdfstring{$\Delta N_{\rm eff}$}{Delta N\_eff}}\label{sec:spin_results}
\begin{figure}[!htbp]
    \centering
    \includegraphics[width=\linewidth]{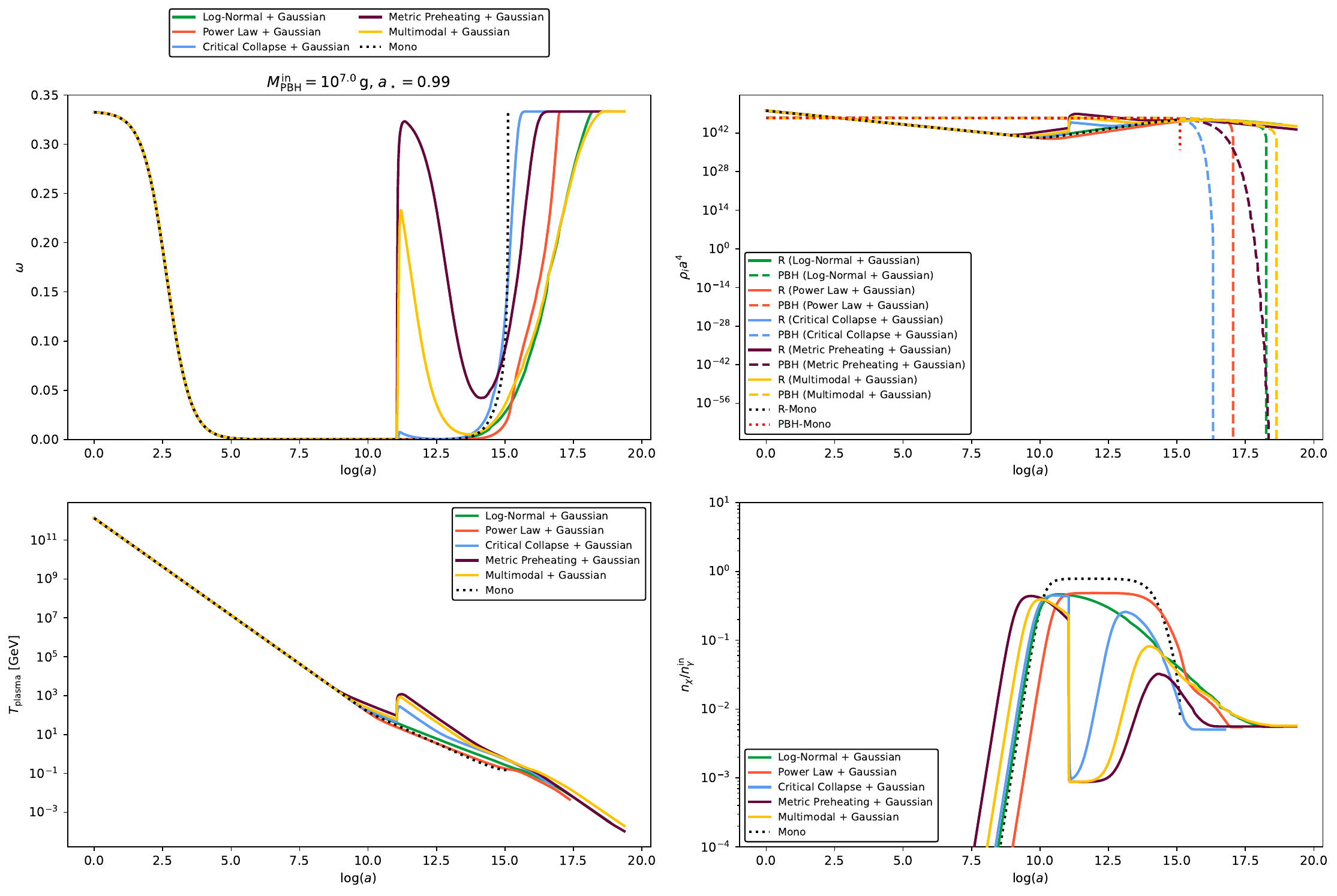}
    \caption{Same as Fig.~\ref{fig:a} but for $a_\star = 0.99$,
with a Gaussian spin distribution of width 
$\sigma_{a_\star} = 0.1$ centered at $a_\star = 0.99$ 
(truncated at $a_\star = 1$ due to the physical bound on 
the spin parameter), for $M_{\rm PBH}^{\rm in} = 10^7\,\mathrm{g}$, 
sDR $= 2.0$, and multimodal results shown for the MW (mean-mass-weighted) scenario 
(see Table~\ref{tab:multimodal_robustness} for the EA and NW variants).
\textit{Top left:} equation-of-state parameter $w$ vs 
$\log(a)$, showing the transition from matter- to 
radiation-domination at PBH evaporation.
\textit{Top right:} comoving SM radiation (solid) and PBH 
(dashed) energy densities $\rho_i a^4$ vs $\log(a)$ for 
each distribution.
\textit{Bottom left:} plasma temperature $T_{\rm plasma}$ 
vs $\log(a)$.
\textit{Bottom right:} dark radiation number density ratio 
$n_\chi/n_\gamma^{\rm in}$ vs $\log(a)$, illustrating the 
enhanced dark radiation production relative to the 
$a_\star = 0.0$ case due to superradiant Hawking emission 
of near-extremal Kerr black holes.}
    \label{fig:b}
\end{figure}

\begin{figure}[!htbp]
    \centering
    \includegraphics[width=\linewidth]{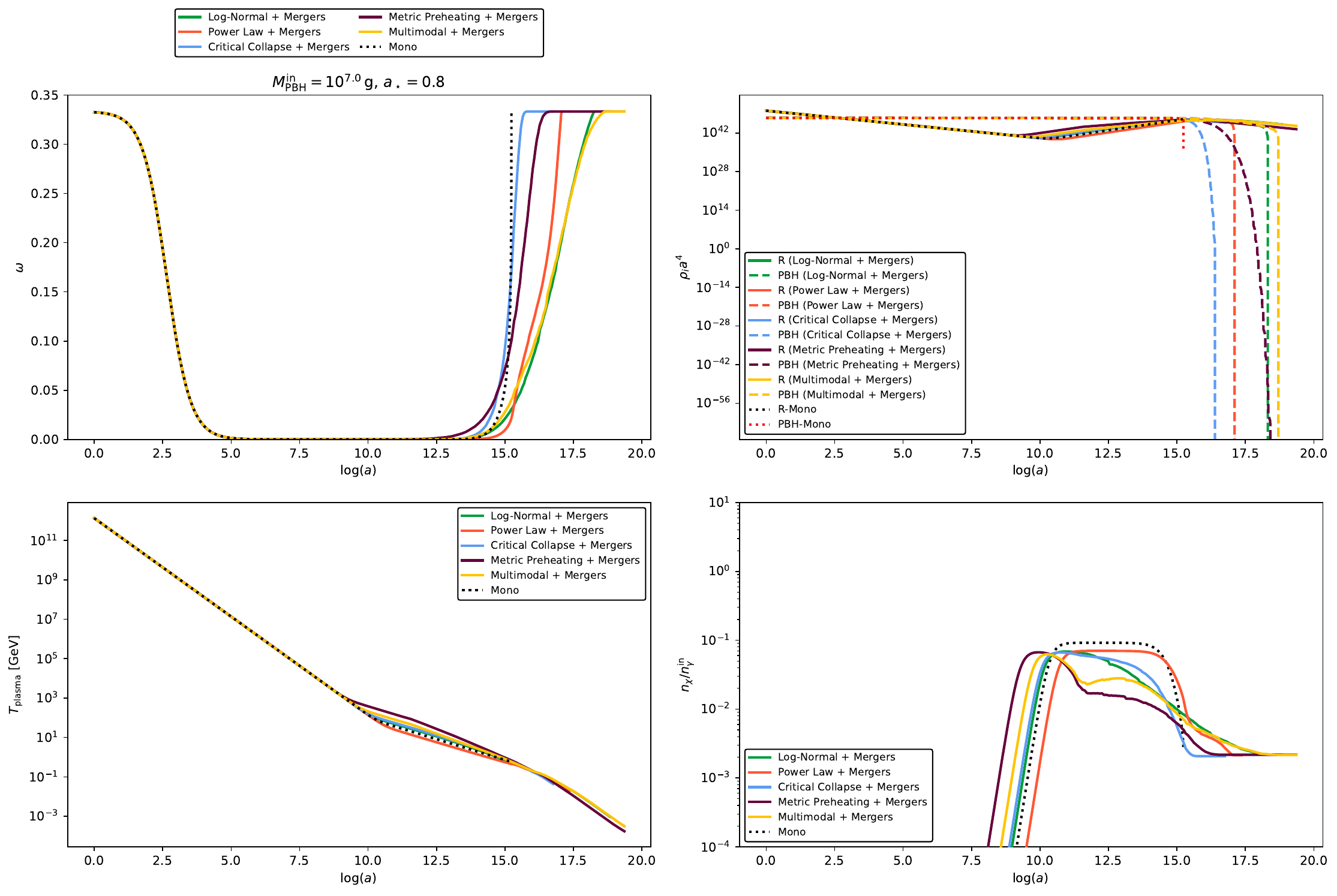}
    \caption{Same as Fig.~\ref{fig:a} but for $a_\star = 0.8$,
with the universal merger-driven spin distribution of 
Fishbach et al.~\cite{fishbach2017ligo}, valid for 
$0.4 \leq a_\star \leq 0.94$, for 
$M_{\rm PBH}^{\rm in} = 10^7\,\mathrm{g}$, sDR $= 2.0$, 
and multimodal results shown for the MW (mean-mass-weighted) scenario.
\textit{Top left:} equation-of-state parameter $w$ vs 
$\log(a)$, showing the transition from matter- to 
radiation-domination at PBH evaporation.
\textit{Top right:} comoving SM radiation (solid) and PBH 
(dashed) energy densities $\rho_i a^4$ vs $\log(a)$ for 
each distribution.
\textit{Bottom left:} plasma temperature $T_{\rm plasma}$ 
vs $\log(a)$.
\textit{Bottom right:} dark radiation number density ratio 
$n_\chi/n_\gamma^{\rm in}$ vs $\log(a)$, illustrating the 
intermediate dark radiation production between the 
$a_\star = 0.0$ and $a_\star = 0.99$ cases, consistent 
with the monotonic dependence of $\Delta N_{\rm eff}$ on spin.}
    \label{fig:c}
\end{figure}

We now examine the role of the initial PBH spin parameter $a^*$ 
on $\Delta N_{\rm eff}$. Tables~\ref{tab:dneff_a099} and~\ref{tab:dneff_a08} 
present results for $a^* = 0.99$ and $0.8$ respectively, while 
Figs.~\ref{fig:b} and~\ref{fig:c} illustrate the spin dependence 
for the full mass and spin distribution cases.

Two physically motivated spin distributions are considered. 
For $a_\star = 0.99$, we adopt a Gaussian profile centered 
at $a_\star$ with width $\sigma_{a_\star} = 0.1$, representing 
PBHs formed with a large initial spin; this distribution is 
truncated at $a_\star = 1$ due to the physical bound on the 
spin parameter. For $a_\star = 0.8$, we adopt the universal 
merger-driven distribution of Fishbach et al.~\cite{fishbach2017ligo}, 
valid for $0.4 \leq a_\star \leq 0.94$, representing PBHs 
that acquired their spin through hierarchical mergers. 
The role of large initial spin in enhancing $N_{\rm eff}$ 
through superradiant Hawking emission has been emphasized 
previously in Ref.~\cite{masina2021dark}; our results confirm 
and extend this finding to the case of extended and 
multimodal mass distributions.

\paragraph*{Quantifying the spin--distribution interplay.}
The mechanism derived in Sec.~\ref{sec:kerr_emission} makes a 
sharp, testable prediction: the sensitivity of $\Delta N_{\rm eff}$ 
to the initial spin $a_\star$ should grow with the spin sDR of the 
emitted dark radiation quantum, since only spin sDR$\geq 1$ channels 
experience appreciable superradiant amplification (Sec.~\ref{sec:kerr_emission}), 
and this sensitivity should be systematically larger for the 
monochromatic benchmark than for extended distributions, because 
the monochromatic population sustains $a_\star \simeq 0.99$ for its 
entire (single) characteristic mass, whereas an extended population 
averages over a spread of formation masses with correspondingly 
different spin-down timescales $\tau_{a_\star}$. Table~\ref{tab:spin_enhancement} 
confirms this directly, reporting the ratio 
$R_i \equiv \Delta N_{\rm eff}^i(a_\star=0.99)/\Delta N_{\rm eff}^i(a_\star=0)$ 
extracted from Tables~\ref{tab:dneff_a0} and~\ref{tab:dneff_a099} for 
each distribution $i$ and each sDR.

\begin{table}[!htbp]
\centering
\footnotesize
\renewcommand{\arraystretch}{1.2}
\setlength{\tabcolsep}{6pt}
\begin{tabular}{|c|c|c|c|c|c|c|}
\hline
\textbf{sDR} & \textbf{Mono} & \textbf{LN} & \textbf{PL} & \textbf{CC} & \textbf{MP} & \textbf{MM} \\
\hline
0.0 & 0.86 & 1.01 & 0.95 & 1.45 & 1.51 & 0.95 \\
0.5 & 0.95 & 1.01 & 0.93 & 1.60 & 1.67 & 1.06 \\
1.0 & 1.24 & 1.05 & 1.03 & 1.96 & 2.16 & 1.11 \\
2.0 & 12.59 & 6.66 & 7.00 & 10.22 & 12.00 & 7.37 \\
\hline
\end{tabular}
\caption{Spin-induced enhancement ratio 
$R_i \equiv \Delta N_{\rm eff}^i(a_\star=0.99)/\Delta N_{\rm eff}^i(a_\star=0)$, 
computed from Tables~\ref{tab:dneff_a0} and~\ref{tab:dneff_a099}, for 
each mass distribution $i$ and dark radiation spin sDR. The sharp 
growth of $R_i$ with sDR, and its systematically larger value for 
the monochromatic and narrow (MP, CC) distributions than for the 
broad (LN, PL) ones at sDR$=2$, is the direct numerical signature 
of the superradiance mechanism derived in Sec.~\ref{sec:kerr_emission}.}
\label{tab:spin_enhancement}
\end{table}

For sDR $\in \{0,0.5\}$, where superradiant amplification is negligible 
or absent (Sec.~\ref{sec:kerr_emission}), the spin-induced enhancement 
$R_i$ remains modest for every distribution ($R_i \lesssim 1.7$), but is 
not uniformly close to unity. The monochromatic, log-normal, power-law, 
and multimodal cases stay within $\sim\!15\%$ of unity 
($R_{\rm mono}\simeq0.86$--$0.95$, $R_{\rm LN}\simeq1.01$, 
$R_{\rm PL}\simeq0.93$--$0.95$, $R_{\rm MM}\simeq0.95$--$1.06$): at these 
sDR values, raising $a_\star$ has essentially no superradiant channel to 
act through, and the residual variation of $R_i$ for these four cases is 
driven by sub-dominant effects of spin on the non-superradiant part of 
the spectrum. The two narrowest distributions, critical collapse and 
metric preheating, depart from this near-unity pattern already at 
sDR $\leq 0.5$, with $R_{\rm CC}\simeq1.45$--$1.60$ and 
$R_{\rm MP}\simeq1.51$--$1.67$ (a $45$--$67\%$ enhancement, the upper 
end of the $R_i\lesssim1.7$ range quoted above). This follows directly from Eq.~\eqref{eq:kerr_temperature}: at fixed $M$, the 
Hawking temperature $T_{\rm H}$ decreases monotonically with $a_\star$, so every 
PBH in the population experiences a lower evaporation temperature as spin 
increases, independently of any superradiant channel. Because CC and MP are the 
most tightly peaked of the five distributions around $M_c$ 
(Sec.~\ref{sec:distributions}), nearly their entire population is exposed to 
essentially the same value of this shift, whereas the broader LN, PL, and MM 
distributions average it over a wide spread of formation masses, diluting its net 
effect on the evaporation history. This distribution-width dependence of a 
temperature effect already present in the Schwarzschild-to-Kerr formalism of 
Sec.~\ref{sec:kerr_emission}, rather than any additional mechanism, accounts for 
the earlier departure of $R_{\rm CC}$ and $R_{\rm MP}$ from unity. For sDR$=1$, the 
onset of superradiance in the vector channel is already visible 
($R_{\rm mono}\simeq1.24$). The effect becomes dramatic for sDR$=2$: 
the monochromatic benchmark is enhanced by more than an order of 
magnitude ($R_{\rm mono}\simeq12.6$), while the extended distributions 
are enhanced by a smaller, though still substantial, factor of 
$6.7$--$12.0$, with the narrower distributions (MP, CC) tracking the 
monochromatic case most closely and the broadest distribution (LN) 
showing the weakest enhancement. This is exactly the pattern expected 
from Eq.~\eqref{eq:dastardt}: because the graviton (spin-2) channel is 
by far the most superradiant~\cite{press1972floating}, only PBHs that 
remain close to $a_\star=0.99$ throughout their evolution can fully 
exploit it, and the monochromatic population -- and, to a lesser 
degree, the narrowly-peaked MP and CC distributions -- do so more 
effectively than the broad LN and PL mass functions, whose lighter 
constituent PBHs spin down (Sec.~\ref{sec:kerr_emission}) well before 
the characteristic evaporation time. The ordering 
$R_{\rm MP}, R_{\rm CC} > R_{\rm MM} > R_{\rm PL}, R_{\rm LN}$ at 
sDR$=2$ is the qualitative inverse of the enhancement hierarchy reported in 
Table~\ref{tab:dneff_a0} for $a_\star=0$: the two narrowest distributions 
(CC, MP), which show the weakest mass-broadening enhancement at $a_\star=0$, 
become the most spin-sensitive at $a_\star=0.99$, while the broadest 
distributions (LN, PL) show the opposite trend. We stress that this is 
a cluster-level inversion (broad vs.\ narrow distributions) rather than 
an exact term-by-term reversal of the five-way ranking: the pairwise 
relation $\mathrm{MP}>\mathrm{CC}$ already present at $a_\star=0$ is 
preserved, not reversed, at sDR$=2$ ($R_{\rm MP}>R_{\rm CC}$), so only 
the broad-versus-narrow grouping -- not the internal order within each 
group -- is inverted. This grouped inversion is nonetheless the direct 
numerical origin of the ratio inversion, 
$\Delta N_{\rm eff}/\Delta N_{\rm eff}^{\rm mono}<1$, reported in 
Table~\ref{tab:dneff_a099} for sDR$=2$ and $a_\star=0.99$.

\begin{table}[!htbp]
\centering
\footnotesize
\setlength{\tabcolsep}{4pt}
\renewcommand{\arraystretch}{1.2}
\resizebox{\linewidth}{!}{%
\begin{tabular}{|c|c|c|
    S[table-format=1.6, round-precision=6]|
    S[table-format=1.6, round-precision=6]|
    S[table-format=1.4, round-precision=4]|
    S[table-format=2.2, round-precision=2]|}
\hline
\textbf{Mass Distribution Type} 
 & \textbf{Free Parameters} & \textbf{sDR} 
 & $\mathbf{\Delta N_{\text{eff}}}$ 
 & $\mathbf{\Delta N_{\text{eff}}^{\text{mono}}}$ 
 & \textbf{Ratio} & \textbf{Diff (\%)} \\
\hline
\multirow{4}{*}{Log-Normal} & \multirow{4}{*}{$\sigma = 1.0$} & 
0.0 & 0.10548717378565992 & 0.048727515297732525 & 2.1648379389163854 & 116.48379389163853 \\
\cline{3-7}
& & 0.5 & 0.11511365325796762 & 0.058211837223142980 & 1.9774956220107494 & 97.74956220107494 \\
\cline{3-7}
& & 1.0 & 0.047967252152023236 & 0.030766027725740365 & 1.5590979953480144 & 55.909799534801444 \\
\cline{3-7}
& & 2.0 & 0.035103377014061480 & 0.036097838534849710 & 0.9724509399689360 & -2.7549060031064108 \\
\hline
\multirow{4}{*}{Power Law} & \multirow{4}{*}{$\sigma=1.0,\ \alpha=2.5$} & 
0.0 & 0.09076548677910345 & 0.048727515297732525 & 1.8627152692788158 & 86.27152692788157 \\
\cline{3-7}
& & 0.5 & 0.09637747609374958 & 0.058211837223142980 & 1.6556336424206397 & 65.56336424206397 \\
\cline{3-7}
& & 1.0 & 0.04317590753548345 & 0.030766027725740365 & 1.4033630834753610 & 40.336308347536104 \\
\cline{3-7}
& & 2.0 & 0.033834232286105895 & 0.036097838534849710 & 0.9372924712221071 & -6.2707528777892865 \\
\hline
\multirow{4}{*}{Critical Collapse} & \multirow{4}{*}{---} & 
0.0 & 0.08554011217812973 & 0.048727515297732525 & 1.7554786377976930 & 75.54786377976929 \\
\cline{3-7}
& & 0.5 & 0.10222493214701955 & 0.058211837223142980 & 1.7560849652479018 & 75.60849652479017 \\
\cline{3-7}
& & 1.0 & 0.05065689093517405 & 0.030766027725740365 & 1.6465203563732090 & 64.65203563732089 \\
\cline{3-7}
& & 2.0 & 0.030301833298817240 & 0.036097838534849710 & 0.8394362246804092 & -16.056377531959075 \\
\hline
\multirow{4}{*}{Metric Preheating} & \multirow{4}{*}{---} & 
0.0 & 0.10060086194964771 & 0.048727515297732525 & 2.0645596504348958 & 106.4559650434896 \\
\cline{3-7}
& & 0.5 & 0.12083555716756195 & 0.058211837223142980 & 2.0757901301820100 & 107.5790130182010 \\
\cline{3-7}
& & 1.0 & 0.06247253885736797 & 0.030766027725740365 & 2.0305688928798693 & 103.0568892879869 \\
\cline{3-7}
& & 2.0 & 0.03960895025730543 & 0.036097838534849710 & 1.0972665363070424 & 9.72665363070425 \\
\hline
\multirow{4}{*}{Multimodal (MW)} & \multirow{4}{*}{\shortstack{$\sigma=1.0,\ \alpha=2.5$\\ 
See Sec.~\ref{sec:multimodal}}} &
0.0 & 0.09579856936089960 & 0.048727515297732525 & 1.9660056289666277 & 96.60056289666277 \\
\cline{3-7}
& & 0.5 & 0.11557618452058847 & 0.058211837223142980 & 1.9854412785075170 & 98.5441278507517 \\
\cline{3-7}
& & 1.0 & 0.04911547037829843 & 0.030766027725740365 & 1.5964189727751568 & 59.64189727751568 \\
\cline{3-7}
& & 2.0 & 0.03736251039356332 & 0.036097838534849710 & 1.0350345591327488 & 3.503455913274876 \\
\hline
\end{tabular}
}
\caption{Excess effective number of relativistic species $\Delta N_{\rm eff}$ 
for each mass distribution compared to the monochromatic benchmark, 
with $\mathbf{M_{\rm PBH}^{\rm in}}=10^7\,\mathrm{g}$ 
and a Gaussian spin distribution centered at $\mathbf{a^{\ast}}=0.99$ 
with width $\sigma_{a_\star}=0.1$. The multimodal result 
uses the mean-mass-weighted (MW) population fractions.
The column ``Free Parameters'' lists the 
parameters varied in each distribution. 
For the critical collapse and metric preheating 
distributions, no free parameters are varied 
(indicated by ---): their shapes are fully 
determined by the physics of gravitational 
collapse~\cite{niemeyer1998near} and metric 
preheating~\cite{cheek2023evaporation, 
martin2020primordial, auclair2021primordial}, 
respectively.
Here sDR denotes the spin of the dark radiation component, 
Ratio $= \Delta N_{\rm eff}/\Delta N_{\rm eff}^{\rm mono}$, 
and Diff $= 100\times(\Delta N_{\rm eff} - \Delta N_{\rm eff}^{\rm mono})/
\Delta N_{\rm eff}^{\rm mono}$. All values are quoted to four significant figures, 
consistent with the numerical integration tolerance 
of FRISHBEE ($\text{rtol} = \text{atol} = 10^{-5}$).}
\label{tab:dneff_a099}
\end{table}

\begin{table}[!htbp]
\centering
\footnotesize
\setlength{\tabcolsep}{4pt}
\renewcommand{\arraystretch}{1.2}
\resizebox{\linewidth}{!}{%
\begin{tabular}{|c|c|c|
    S[table-format=1.6, round-precision=6]|
    S[table-format=1.6, round-precision=6]|
    S[table-format=1.4, round-precision=4]|
    S[table-format=2.2, round-precision=2]|}
\hline
\textbf{Mass Distribution Type} 
 & \textbf{Free Parameters} & \textbf{sDR} 
 & $\mathbf{\Delta N_{\text{eff}}}$ 
 & $\mathbf{\Delta N_{\text{eff}}^{\text{mono}}}$ 
 & \textbf{Ratio} & \textbf{Diff (\%)} \\
\hline
\multirow{4}{*}{Log-Normal} & \multirow{4}{*}{$\sigma = 1.0$} & 
0.0 & 0.09953601319774114 & 0.05129478603631756 & 1.9404703847924814 & 94.04703847924812 \\
\cline{3-7}
& & 0.5 & 0.11458187725097853 & 0.059413941518340084 & 1.9285351943130895 & 92.85351943130894 \\
\cline{3-7}
& & 1.0 & 0.05282754942992810 & 0.027819216141979308 & 1.8989589483871590 & 89.89589483871589 \\
\cline{3-7}
& & 2.0 & 0.013588386945641980 & 0.008142225325165502 & 1.6688787650771355 & 66.88787650771354 \\
\hline
\multirow{4}{*}{Power Law} & \multirow{4}{*}{$\sigma=1.0,\ \alpha=2.5$} & 
0.0 & 0.09508978946015095 & 0.05129478603631756 & 1.8537905469149594 & 85.37905469149592 \\
\cline{3-7}
& & 0.5 & 0.10989259255734132 & 0.059413941518340084 & 1.8496095318540570 & 84.96095318540570 \\
\cline{3-7}
& & 1.0 & 0.05033827572797141 & 0.027819216141979308 & 1.8094785802397484 & 80.94785802397485 \\
\cline{3-7}
& & 2.0 & 0.012974339739795834 & 0.008142225325165502 & 1.5934636075097950 & 59.34636075097949 \\
\hline
\multirow{4}{*}{Critical Collapse} & \multirow{4}{*} {---} & 
0.0 & 0.08891560216223976 & 0.05129478603631756 & 1.7334237850078180 & 73.34237850078180 \\
\cline{3-7}
& & 0.5 & 0.09244007257048360 & 0.059413941518340084 & 1.5558650075748452 & 55.58650075748452 \\
\cline{3-7}
& & 1.0 & 0.04726329605236429 & 0.027819216141979308 & 1.6989442050110029 & 69.89442050110027 \\
\cline{3-7}
& & 2.0 & 0.011734755607907287 & 0.008142225325165502 & 1.4412221646136723 & 44.12221646136724 \\
\hline
\multirow{4}{*}{Metric Preheating} & \multirow{4}{*} {---} & 
0.0 & 0.10473506582913261 & 0.05129478603631756 & 2.0418267415128400 & 104.18267415128400 \\
\cline{3-7}
& & 0.5 & 0.12178431024798855 & 0.059413941518340084 & 2.0497598229600000 & 104.97598229599998 \\
\cline{3-7}
& & 1.0 & 0.05424794481270636 & 0.027819216141979308 & 1.9500170147082612 & 95.00170147082612 \\
\cline{3-7}
& & 2.0 & 0.014420916224007890 & 0.008142225325165502 & 1.7711271363906604 & 77.11271363906604 \\
\hline
\multirow{4}{*}{Multimodal (MW)} & \multirow{4}{*}{\shortstack{$\sigma=1.0,\ \alpha=2.5$\\ 
See Sec.~\ref{sec:multimodal}}} &
0.0 & 0.10091565472294217 & 0.05129478603631756 & 1.9673667154297556 & 96.73667154297556 \\
\cline{3-7}
& & 0.5 & 0.11676070105877280 & 0.059413941518340084 & 1.9652071226873702 & 96.52071226873701 \\
\cline{3-7}
& & 1.0 & 0.05267143965999887 & 0.027819216141979308 & 1.8933473679194526 & 89.33473679194526 \\
\cline{3-7}
& & 2.0 & 0.013696758577387330 & 0.008142225325165502 & 1.6821885946897355 & 68.21885946897356 \\
\hline
\end{tabular}
}
\caption{Excess effective number of relativistic species $\Delta N_{\rm eff}$ 
for each mass distribution compared to the monochromatic benchmark, 
with $\mathbf{M_{\rm PBH}^{\rm in}}=10^7\,\mathrm{g}$ and a spin distribution 
from hierarchical mergers~\cite{fishbach2017ligo} (mean $\langle a_\star \rangle \approx 0.8$).
The multimodal result uses the mean-mass-weighted (MW) population fractions.
The column ``Free Parameters'' lists the 
parameters varied in each distribution. 
For the critical collapse and metric preheating 
distributions, no free parameters are varied 
(indicated by ---): their shapes are fully 
determined by the physics of gravitational 
collapse~\cite{niemeyer1998near} and metric 
preheating~\cite{cheek2023evaporation, 
martin2020primordial, auclair2021primordial}, 
respectively.
Here sDR denotes the spin of the dark radiation component, 
Ratio $= \Delta N_{\rm eff}/\Delta N_{\rm eff}^{\rm mono}$, 
and Diff $= 100\times(\Delta N_{\rm eff} - \Delta N_{\rm eff}^{\rm mono})/
\Delta N_{\rm eff}^{\rm mono}$. All values are quoted to four significant figures, 
consistent with the numerical integration tolerance 
of FRISHBEE ($\text{rtol} = \text{atol} = 10^{-5}$).}
\label{tab:dneff_a08}
\end{table}

Increasing the initial spin parameter from $a^* = 0.0$ to $a^* = 0.99$ enhances $\Delta N_{\rm eff}$ for all mass distributions, but the size of the enhancement depends strongly on both the distribution shape and the spin sDR of the dark radiation species, as quantified by the ratio $R_i$ of Table~\ref{tab:spin_enhancement} and traced back to the superradiance mechanism of Sec.~\ref{sec:kerr_emission}. For sDR $\leq 0.5$, where superradiant amplification is negligible, the spin-induced enhancement stays modest for every distribution ($R_i \lesssim 1.7$). The effect grows sharply with sDR: at sDR$=2$, the critical collapse and metric preheating distributions, being the narrowest, exhibit the largest enhancement (a factor of $10$--$12$), the broader log-normal and power-law distributions show a more modest increase (a factor of $\sim 6.7$--$7.0$), and the multimodal distribution falls in between (a factor of $\sim 7.4$). This pattern is consistent with the well-known result that Kerr black holes evaporate faster than Schwarzschild black holes of the same mass~\cite{PhysRevD.13.198, page1976particle}: the superradiant amplification of Hawking emission for rotating black holes increases the total power radiated, an effect that grows with the spin of the emitted quantum (Sec.~\ref{sec:kerr_emission}). However, the magnitude of the enhancement is diluted in broader mass distributions, where a significant fraction of PBHs have masses far from the characteristic scale $10^7\,\mathrm{g}$, spin down (Sec.~\ref{sec:kerr_emission}) well before the characteristic evaporation time, and are therefore less sensitive to the spin-enhanced evaporation.
The spin-induced enhancement found here is 
consistent with and extends the single-distribution 
results of Masina~\cite{masina2021dark}, who first 
identified the significant increase in $N_{\rm eff}$ 
from near-extremal Kerr PBHs; our analysis 
demonstrates that this enhancement is modulated 
by the width of the mass distribution in a 
quantifiable way.

The intermediate case $a^* = 0.8$ yields enhancements consistent 
with a smooth interpolation between the Schwarzschild and 
near-extremal Kerr limits, confirming the monotonic dependence 
of $\Delta N_{\rm eff}$ on spin. The monochromatic benchmark 
$\Delta N_{\rm eff}^{\rm mono}$ also increases with spin, so 
the ratio $\Delta N_{\rm eff}/\Delta N_{\rm eff}^{\rm mono}$ 
remains approximately constant across spin values for each 
distribution. This implies that the relative enhancement due to 
extended distributions is largely independent of spin, and the 
two effects, namely mass distribution broadening and spin, factorize 
to good approximation.

The nearly universal spin distributions from hierarchical 
mergers, peaked near $a_\star \approx 0.7$~\cite{fishbach2017ligo}, 
would therefore produce $\Delta N_{\rm eff}$ enhancements 
intermediate between the $a^* = 0.0$ and $a^* = 0.99$ cases, 
with the mass distribution shape remaining the dominant driver 
of the enhancement.

\begin{figure}[!htbp]
\centering
\begin{minipage}{.72\linewidth}
\centering
\includegraphics[width=\linewidth]{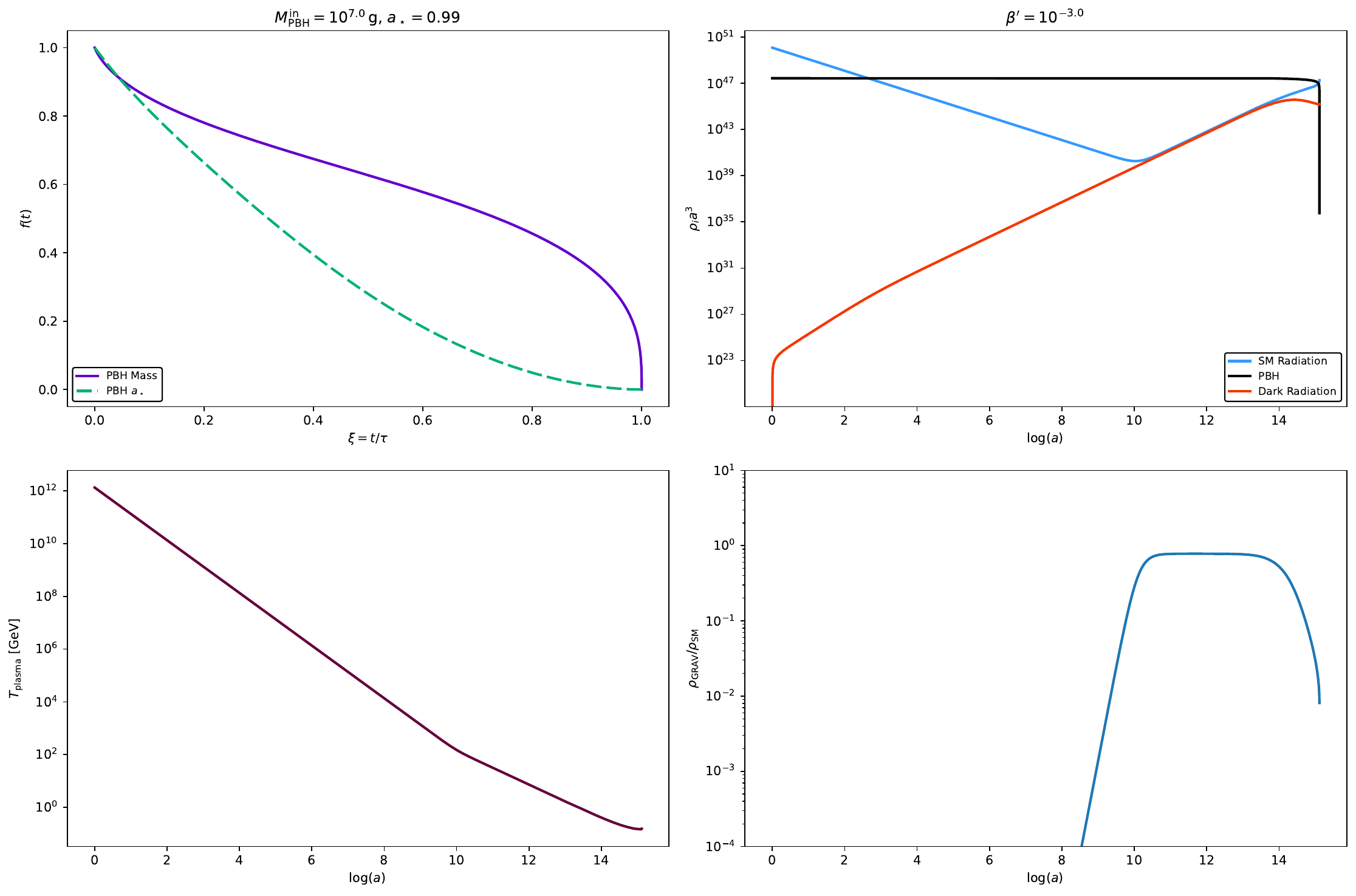}
\smallskip
\text{(a)}
\end{minipage}
\medskip
\begin{minipage}{.72\linewidth}
\centering
\includegraphics[width=\linewidth]{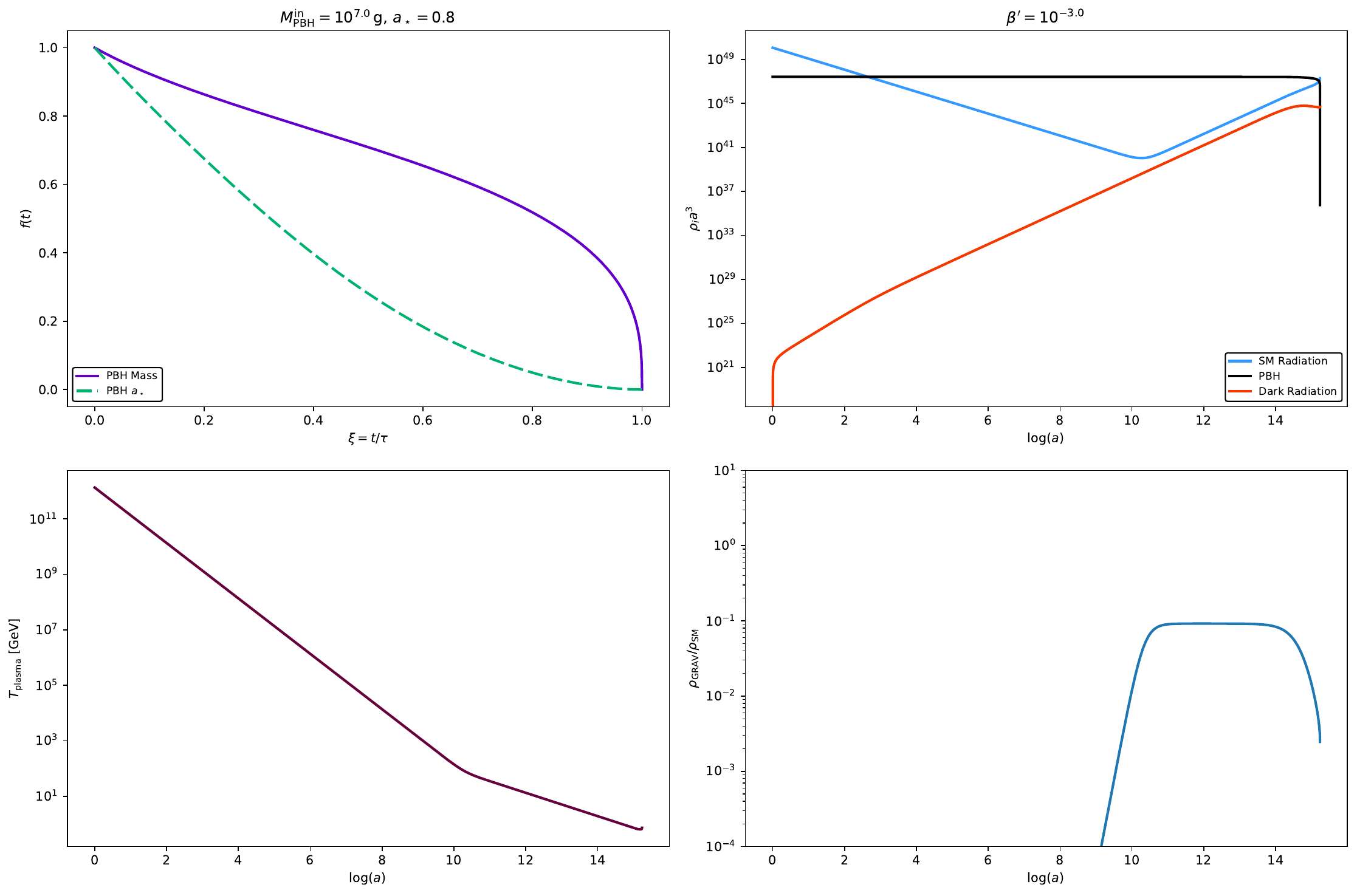}
\smallskip
\text{(b)}
\end{minipage}
\caption{Cosmological evolution for the monochromatic PBH 
scenario with $M_{\rm PBH}^{\rm in} = 10^7\,\mathrm{g}$, 
$\beta' = 10^{-3}$, and sDR $= 2.0$, for two spin values:
(a) $a_\star = 0.99$ and (b) $a_\star = 0.8$.
\textit{Top left:} normalized mass $f(t) = M(t)/M_{\rm PBH}^{\rm in}$ 
(solid purple) and spin parameter $a_\star(t)$ (dashed green) 
vs normalized time $\xi = t/\tau$, showing that faster-spinning 
PBHs lose their angular momentum more rapidly during evaporation.
\textit{Top right:} comoving energy densities $\rho_i a^3$ 
for SM radiation (blue), PBH (black), and dark radiation 
(orange) vs $\log(a)$.
\textit{Bottom left:} plasma temperature $T_{\rm plasma}$ 
vs $\log(a)$.
\textit{Bottom right:} dark-to-SM radiation ratio 
$\rho_{\rm GRAV}/\rho_{\rm SM}$ vs $\log(a)$, showing 
that the near-extremal case $a_\star = 0.99$ produces 
a significantly larger dark radiation fraction than 
$a_\star = 0.8$, consistent with the enhanced Hawking 
emission of rotating black holes.}
\label{fig:Mono}
\end{figure}

An exception to the systematic enhancement 
pattern arises for sDR $= 2.0$ at $a_\star = 0.99$, 
where the log-normal, power-law, and critical 
collapse distributions yield ratios 
$\Delta N_{\rm eff}/\Delta N_{\rm eff}^{\rm mono} < 1$ 
(see Table~\ref{tab:dneff_a099}). This inversion 
reflects a competition between two effects. 
On one hand, extended distributions generically 
enhance $\Delta N_{\rm eff}$ through the contribution 
of light PBHs that evaporate at high temperatures. 
On the other hand, for near-extremal spins and 
spin-2 dark radiation, the Hawking emission rate 
is strongly amplified by superradiance, and this 
amplification is maximally sustained in the 
monochromatic case where all PBHs maintain 
$a_\star = 0.99$ throughout their evaporation. 
In extended distributions, lighter PBHs lose their 
angular momentum more rapidly during evaporation 
(as illustrated in Fig.~\ref{fig:Mono} and quantified 
via the spin-down timescale $\tau_{a_\star}$ of 
Sec.~\ref{sec:kerr_emission}), reducing 
their time-averaged spin and hence their 
superradiant contribution to sDR $= 2$ emission. 
When this spin-loss effect outweighs the 
mass-broadening enhancement, which occurs 
specifically for spin-2 dark radiation at near-extremal 
spin, the monochromatic approximation 
overestimates $\Delta N_{\rm eff}$, as quantified 
in Table~\ref{tab:spin_enhancement}. The metric 
preheating and multimodal distributions, being 
more narrowly peaked around $M_c$, are less 
susceptible to this effect and retain ratios 
above unity. This result demonstrates that the 
sign of the monochromatic approximation error 
is not universal but depends on the interplay 
between the mass distribution shape, the initial 
spin, and the spin of the dark radiation species.

The merger-driven spin distribution for $a_\star = 0.8$ 
(Fig.~\ref{fig:c}) produces qualitatively similar 
enhancement patterns to the Gaussian case at $a_\star = 0.99$ 
(Fig.~\ref{fig:b}), with the metric preheating distribution 
consistently yielding the largest enhancement among all 
mass distributions.

Fig.~\ref{fig:Mono} shows the cosmological evolution for the 
monochromatic case at $a^* = 0.99$ (panel a) and $a^* = 0.8$ 
(panel b). The dark-to-SM radiation ratio 
$\rho_{\rm GRAV}/\rho_{\rm SM}$ is significantly larger for 
$a_\star = 0.99$ than for $a_\star = 0.8$, confirming the 
spin dependence of the dark radiation production that 
underlies the $\Delta N_{\rm eff}$ enhancements reported 
in Tables~\ref{tab:dneff_a099} and~\ref{tab:dneff_a08}.

\section{Conclusion}\label{sec:conclusion}

We have gone beyond the monochromatic approximation for PBH Hawking evaporation, studying the imprint on $\Delta N_{\rm eff}$ of realistic extended and, for the first time, multimodal PBH mass and spin distributions. Using \texttt{FRISHBEE}, our extension of the public \texttt{FRISBHEE} code, we computed $\Delta N_{\rm eff}$ for four physically motivated mass functions (log-normal, power-law, critical collapse, metric preheating) and for a multimodal mixture of all four, evaluated across three characteristic mass scales, three spin configurations, and four dark-radiation spins.

The central methodological advance of this work, relative to Refs.~\cite{lunardini2020dirac, masina2021dark, hasegawa2019mev, carr2021constraints, cheek2023evaporation}, is that the multimodal population fractions $\alpha_i$ are not free parameters fitted to the data: they follow self-consistently from the primordial collapse probability, $\beta'_i \simeq \gamma_i\,\mathrm{erfc}\!\left(\dfrac{\delta_c}{\sqrt{2}\,\sigma(k_i)}\right)$, which ties the multimodal mass function directly to the amplitude of the primordial curvature power spectrum at each formation scale. This closes the gap between multi-feature inflationary model building and the $N_{\rm eff}$ observable: for any specific inflationary model with several power-spectrum features, the $\alpha_i$ -- and hence the predicted $\Delta N_{\rm eff}$ -- are, in principle, calculable rather than assumed. We tested this framework against three benchmark weighting scenarios (equal abundance, mean-mass weighted, number-density weighted) and found the resulting enhancement robust in all three (Table~\ref{tab:multimodal_robustness}), confirming that the multimodal signal is a genuine prediction of multi-channel PBH formation rather than an artifact of a particular choice of weights.

Numerically, extended mass functions enhance $\Delta N_{\rm eff}$ over the monochromatic benchmark by factors from $\sim1.03$ (critical collapse) to $\sim1.84$ (log-normal), with the power-law ($\sim1.69$) and metric-preheating ($\sim1.15$) distributions in between, and the multimodal mixture ranging from $\sim1.43$ (equal-abundance weighting) to $\sim1.77$ (mean-mass weighting). This hierarchy, $LN>MM>PL>MP>CC$, and the underlying dominance of the shape of the mass function over its characteristic mass, is preserved across the full pre-BBN window $M_{\rm PBH}^{\rm in}\in\{10^5,10^7,10^8\}\,\mathrm{g}$ (Table~\ref{tab:stability}).

These enhancement factors are not merely of academic interest: they determine whether the PBH evaporation signal is within reach of upcoming CMB experiments. At $M_{\rm PBH}^{\rm in}=10^{7}\,\mathrm{g}$ and scalar dark radiation, the monochromatic prediction, $\Delta N_{\rm eff}^{\rm mono}\simeq0.057$, falls below the projected CMB-S4 sensitivity and is effectively invisible, extending quantitatively the general finding of Ref.~\cite{lunardini2020dirac} that monochromatic PBH populations cannot generate a large $\Delta N_{\rm eff}$. The log-normal, power-law, and multimodal distributions, by contrast, reach $\Delta N_{\rm eff}\simeq0.096$--$0.105$, at the $\sim1.6$--$1.7\sigma$ level for CMB-S4 -- a qualitative, not merely quantitative, difference in observational reach. In other words, whether the PBH evaporation signal is detectable at all by next-generation CMB experiments can depend on the formation channel, making the shape of the PBH mass function -- and, through the multimodal fractions, the shape of the primordial power spectrum -- directly testable by precision measurements of $N_{\rm eff}$.

The role of PBH spin is comparably significant, but its sign depends on the spin of the dark-radiation species itself: spin-induced enhancement is modest for sDR$\leq0.5$ ($R_i\lesssim1.7$ for all distributions, Table~\ref{tab:spin_enhancement}), but grows to a factor of $6.7$--$7.4$ (broad distributions) or $10$--$12$ (narrow distributions) at sDR$=2$, consistent with the increased Hawking emission rate of rotating black holes~\cite{PhysRevD.13.198, page1976particle} and with the spin-induced $N_{\rm eff}$ enhancement first identified by Masina~\cite{masina2021dark}. For near-extremal spin and spin-2 dark radiation specifically, the same superradiant mechanism that boosts the monochromatic signal is diluted by mass averaging in extended distributions, to the point of inverting the enhancement ratio below unity (Table~\ref{tab:dneff_a099}): the monochromatic approximation, which underestimates $\Delta N_{\rm eff}$ at $a_\star=0$, can \emph{overestimate} it at $a_\star=0.99$. This sign-dependence on distribution width is itself a testable prediction, and underlines that neither the monochromatic nor any single extended distribution can be assumed a priori to bound the true signal from either side.

All results reported here are restricted to the pre-BBN window $M_{\rm PBH}^{\rm in}\in\{10^5,10^7,10^8\}\,\mathrm{g}$, where $T_{\rm evap}\gg T_{\rm dec}$ and the electromagnetic dilution of $N_{\rm eff}$ discussed in Sec.~\ref{sec:neff_sign} is absent. Extended distributions with a broad high-mass tail approaching $M_{\rm PBH}\sim10^{9}\,\mathrm{g}$ can partially probe this dilution regime even when centred well below it; a dedicated study of the resulting partial sign change for broad distributions straddling the BBN boundary, and of the companion dark matter relic abundance across the same distributions, is left to forthcoming work. We have focused exclusively on $\Delta N_{\rm eff}$ as the primary cosmological observable sourced by Hawking evaporation in this mass window; other potential imprints of PBH evaporation -- such as modifications to BBN light-element abundances, the generation of a baryon asymmetry, or contributions to the stochastic gravitational-wave background -- are likewise left for future work.

Taken together, these results show that the shape of the PBH mass and spin distributions is not a secondary detail but a primary driver of both the magnitude and the observability of PBH Hawking evaporation. Because our multimodal framework connects the mass-function shape directly to the primordial power spectrum via the collapse probability, precision measurements of $N_{\rm eff}$ by CMB-S4 and the Simons Observatory offer more than a bound on the existence of PBHs: they offer, in principle, a probe of the number and relative strength of the features that seeded them during inflation. \texttt{FRISHBEE}, made public alongside this work, provides the tool needed to confront this class of models with those forthcoming data.

\section*{Data Availability Statement} This manuscript has no associated data.
All data generated during this study (numerical
results for the effective number of relativistic species, $\Delta N_{\rm eff}$ ) are pre-
sented within the article and are fully reproducible using the FRISHBEE
code, publicly available on Zenodo at \url{https://doi.org/10.5281/zenodo.22699162}. No external datasets were used.
\section*{Code Availability Statement} This code/software will be made avail-
able on reasonable request. The code (FRISHBEE)
developed for and used in this study is openly available on Zenodo at
\url{https://doi.org/10.5281/zenodo.22699162.}
\bibliographystyle{spmpsci}
\bibliography{references}
\end{document}